%% file: main.tex
\documentclass[preprint,12pt,authoryear]{elsarticle}

\usepackage{amsmath,amssymb,bm}
\usepackage{graphicx}
\usepackage{booktabs}
\usepackage{multirow}
\usepackage{algorithm}
\usepackage{algpseudocode}
\usepackage[hidelinks]{hyperref}
\usepackage{xcolor}
\usepackage[margin=2.5cm]{geometry}

\newcommand{\E}{\mathbb{E}}
\newcommand{\Var}{\operatorname{Var}}
\newcommand{\Cov}{\operatorname{Cov}}
\newcommand{\tr}{\operatorname{tr}}
\newcommand{\diag}{\operatorname{diag}}
\newcommand{\bp}{\bm{p}}
\newcommand{\bs}{\bm{s}}
\newcommand{\bZ}{\bm{Z}}
\newcommand{\bc}{\bm{c}}
\newcommand{\Dp}{\bm{D}_{\bp}}
\newcommand{\SigN}{\bm{\Sigma}_N}
\newcommand{\dbar}{\bar{\delta}_N}
\newcommand{\Neff}{N_{\mathrm{eff}}}
\newcommand{\one}{\bm{1}}
\newcommand{\He}{\mathrm{He}}

\journal{Computational Statistics \& Data Analysis}

\begin{document}

\begin{frontmatter}

\title{Calibration scoring under dependent seed variables in Cooke's Classical Model: a Mehler-Hermite design-effect correction}

\author{Yuri Calleo\corref{cor1}}
\ead{yuri.calleo@unimercatorum.it}
\cortext[cor1]{Corresponding author.}
\affiliation{organization={Department of Economics, Statistics and Business, Universitas Mercatorum},
            addressline={Piazza Mattei 10},
            postcode={00186},
            city={Rome},
            country={Italy}}

\begin{abstract}
Cooke's Classical Model scores the statistical accuracy of experts by the $p$-value of a likelihood ratio test that compares the bin frequencies of realized seed variables with nominal probabilities, using a $\chi^2$ reference. This calibration score drives performance-based weights and implicitly assumes independent seed variables, although seeds often share a common driver, such as the same indicator in several regions or periods. Under dependence the statistic converges to a weighted sum of $\chi^2_1$ variables, and positive dependence makes the classical score anti-conservative. Under a Gaussian-copula working model, a Mehler-Hermite expansion gives the long-run covariance of the bin indicators in closed form, defines a design effect and an effective number of seed variables, and shows that the first-order correction coincides with the ad hoc calibration power of existing software. Rao-Scott-type and parametric-bootstrap corrected scores are proposed, together with a restricted maximum likelihood estimator of the nuisance dependence, computed from rank normal scores and invariant to location-scale miscalibration, and a prepivoted bootstrap that propagates its uncertainty. In simulations the classical score rejects up to 68\% of calibrated experts at a nominal 5\% level, whereas the corrected scores remove most of the excess rejection. In 45 published studies, design effects of 1.25-2 shift more than 5\% of the optimized weight in 29-44\% of studies. In three social-science studies with structured seeds, the baseline estimates imply that 18-21 seed variables carry the information of only 11-13 independent ones, although conclusions about individual experts depend on how the dependence is specified.
\end{abstract}


\begin{keyword}
Structured expert judgment \sep Classical Model \sep Calibration score \sep Design effect \sep Hermite polynomials \sep Parametric bootstrap
\end{keyword}

\end{frontmatter}
\section{Introduction}\label{sec:intro}

Structured expert judgment (SEJ) is the standard route to uncertainty quantification when data are scarce, the questions concern the future, or the system is too complex to model from first principles. Typical applications include the ice-sheet contribution to sea-level rise, volcanic hazard, food-borne disease burden, and the reliability of novel engineering systems \citep{aspinall2010route,bamber2013expert,colson2018expert,hanea2021expert}. Among the protocols in use, Cooke's \emph{Classical Model} \citep{cooke1991experts} is distinctive because it is empirically controlled. Before giving their judgments on the target variables, experts assess a set of seed (calibration) variables whose true values are known to the analyst but not to the experts. For each seed, the expert states a few quantiles, usually the 5\%, 50\% and 95\%. The realized values are then used to score every expert on statistical accuracy (calibration), which asks whether the realizations fall in the inter-quantile bins with the stated probabilities, and on informativeness, which measures how concentrated the expert's distributions are. The product of the two scores defines performance-based weights, which are used to build a linear opinion pool, the decision maker (DM) \citep{cooke2000procedures,cooke2008tu}.

The calibration score is the $p$-value of a likelihood-ratio test of the hypothesis that the expert's bin probabilities are correct. If $\bs_e$ is the vector of empirical bin frequencies of expert $e$ over $N$ seeds and $\bp$ is the vector of nominal bin probabilities, the statistic $2N\,I(\bs_e;\bp)$, where $I$ denotes the Kullback-Leibler divergence, is referred to a $\chi^2$ distribution with $K-1$ degrees of freedom, $K$ being the number of bins \citep{cooke1991experts}. This reference distribution is justified only asymptotically, and only if the bin indicators are independent across seeds, that is, if $N\bs_e$ is multinomial. The assumption is seldom stated and rarely checked, although it is often implausible and seed variables are usually chosen from the same measurement campaign as the targets, for good reasons of relevance \citep{cooke2000procedures}. They are therefore often the same indicator in several regions, the same quantity at successive dates, or the same economic aggregate for different groups of countries. If the common driver of such variables happens to deviate from what the expert community expected, every expert's realizations shift towards the same bins together. A perfectly calibrated expert can then receive a very low calibration score without being wrong about anything except one shared anomaly.

Three strands of literature come close to this problem without treating it. First, the Classical Model community has studied the number of seed variables, since the calibration score becomes more discriminating as $N$ grows \citep{eggstaff2014effect,colson2017cross,marti2021performance}. The software implementations offer a ``calibration power'' that rescales the effective number of seed variables. It is presented as a pragmatic device to counteract very low scores when there are many seeds, and its use to steer results is discouraged \citep{cooke2000procedures}; however, no statistical rationale for choosing its value has been given. Second, critics of the Classical Model have pointed out that the calibration score is not a proper scoring rule and that the weights are sensitive to the choice of seeds \citep{clemen2008comment,flandoli2011comparison}, but the dependence structure of the seeds has not been part of that discussion. Third, in forecast verification, it is known that serial and spatial correlation inflate the variability of rank histograms and invalidate $\chi^2$ tests of their flatness \citep{hamill2001interpretation,wilks2011reliability}. In survey statistics, \citet{rao1981analysis,rao1984chi} corrected Pearson and likelihood-ratio tests for complex sampling designs, and \citet{noonan2026exceedance} recently showed that coverage at a single threshold has its own design effect under clustering. None of these results address the multinomial bin structure of the Classical Model, its use of the $p$-value as a weight rather than as a test, or the identifiability problem created by the fact that all experts face the same realized values.

This paper fills that gap. We derive the asymptotic null distribution of the calibration statistic under dependent seeds, which is a weighted sum of $K-1$ independent $\chi^2_1$ variables, and we show that positive dependence makes the classical score anti-conservative. Under a Gaussian-copula working model for the probability integral transforms (PITs) of the realizations, a Mehler-Hermite expansion gives the long-run covariance of the bin indicators in closed form. It separates the geometry of the seed set from the elicitation format and defines a design effect of the seed set and an effective number of seed variables, with explicit bounds and a ceiling under infill sampling. The first-order correction turns out to be identical to the calibration power of existing software, which thereby acquires a precise meaning and becomes estimable. We propose Rao-Scott-type, exact-limit and parametric-bootstrap corrected scores, together with a REML estimator of the nuisance dependence, computed from rank normal scores, that is exactly invariant to location-scale miscalibration, and a prepivoted bootstrap that propagates the uncertainty of that estimate. Seeds need not be geographical: the method applies to any ``question space'' in which a distance between seeds can be defined, for example from shared region, period or quantity type.

The paper combines theory with two kinds of evidence. Simulation studies quantify size, power, robustness to non-Gaussian dependence and the consequences for performance weights. An analysis of 45 published studies from the TU Delft expert judgment database, distributed with the open-source Anduryl software \citep{thart2019update,rongen2020update}, quantifies how sensitive real calibration scores and weights are to the design effect. Three case studies in the social sciences, on political violence, on terrorism and conflict, and on oil markets and economic growth, whose seeds carry explicit region, period and quantity attributes, estimate the dependence from the data, recompute the scores and weights, and examine how sensitive the results are to the specification of the dependence.

The paper is organized as follows. Section~\ref{sec:cm} reviews the Classical Model and makes the independence assumption explicit. Section~\ref{sec:theory} develops the distribution theory and the corrected scores, and Section~\ref{sec:estimation} the estimation of the dependence and the computational procedure. Section~\ref{sec:sim} reports the simulation studies and Section~\ref{sec:app} the analysis of real data. Section~\ref{sec:discussion} discusses implications and limitations and concludes. Derivations are collected in \ref{app:proofs} and computational details in \ref{app:comp}.

\section{The Classical Model and its implicit independence assumption}\label{sec:cm}

\subsection{Elicitation format and calibration score}

Let $e=1,\dots,E$ index experts and $i=1,\dots,N$ index seed variables with realizations $x_1,\dots,x_N$. Each seed is attached to a point $\bm u_i$ of a set $\mathcal{D}$ equipped with a distance; this may be geographical space, time, or a space of question attributes. Expert $e$ states quantiles $q_{ei1}<\dots<q_{ei,K-1}$ of her subjective distribution $F_{ei}$ for $X_i$ at levels $0<\pi_1<\dots<\pi_{K-1}<1$. With $\pi_0=0$ and $\pi_K=1$, the bins $B_{eik}=(q_{ei,k-1},q_{eik}]$, where $q_{ei0}=-\infty$ and $q_{eiK}=+\infty$, have nominal probabilities $p_k=\pi_k-\pi_{k-1}$, $k=1,\dots,K$. We write $\bp=(p_1,\dots,p_K)^\top$ and $\Dp=\diag(\bp)$. In the standard format $K=4$ and $\bp=(0.05,0.45,0.45,0.05)^\top$.

Let $\bZ_{ei}\in\{0,1\}^K$ indicate the bin containing $x_i$, and let $\bs_e=N^{-1}\sum_{i=1}^N\bZ_{ei}$ be the vector of empirical frequencies. The calibration statistic and the calibration score are
\begin{equation}\label{eq:T}
T_e = 2N\,I(\bs_e;\bp) = 2N\sum_{k=1}^K s_{ek}\log\frac{s_{ek}}{p_k},\qquad
C_e^{\mathrm{cl}} = 1-F_{\chi^2_{K-1}}(T_e),
\end{equation}
with $0\log 0=0$. The score $C_e^{\mathrm{cl}}$ is the $p$-value of the likelihood-ratio (G-) test of
\begin{equation}\label{eq:H0}
H_0^{(e)}:\quad \Pr(X_i\in B_{eik})=p_k,\qquad i=1,\dots,N,\;k=1,\dots,K,
\end{equation}
against unrestricted bin probabilities \citep{cooke1991experts,cressie1984multinomial}.

\subsection{Information, weights and the decision maker}

For each seed the analyst fixes an intrinsic range $[L_i,U_i]$, namely the smallest interval containing all experts' quantiles and the realization, extended by 10\% on each side \citep{cooke2000procedures}. Expert $e$'s minimum-information distribution for $X_i$ is piecewise uniform, with mass $p_k$ spread uniformly over each bin truncated to $[L_i,U_i]$. The information score is the mean relative information with respect to the uniform background,
\begin{equation}\label{eq:info}
\bar{I}_e = \frac1N\sum_{i=1}^N\sum_{k=1}^K p_k\log\frac{p_k}{r_{eik}},\qquad r_{eik}=\frac{q_{eik}-q_{ei,k-1}}{U_i-L_i},
\end{equation}
where now $q_{ei0}=L_i$ and $q_{eiK}=U_i$. Global performance weights are
\begin{equation}\label{eq:weights}
w_e(\alpha) \propto C_e\,\bar{I}_e\,\mathbb{1}\{C_e\ge\alpha\},
\end{equation}
and the DM is the linear pool $F_{\mathrm{DM},i}=\sum_e w_e(\alpha)F_{ei}$. The cut-off $\alpha$ is chosen to maximize the DM's own combined score $C_{\mathrm{DM}}\bar I_{\mathrm{DM}}$, with the DM scored as a virtual expert on the seeds \citep{cooke1991experts,cooke2008tu}. The weights and the cut-off therefore depend on the numerical values of the calibration scores and not only on their ranking.

\subsection{Where independence enters}

Under $H_0^{(e)}$ each $\bZ_{ei}$ is marginally multinomial$(1,\bp)$. The step from~\eqref{eq:H0} to the $\chi^2_{K-1}$ reference in~\eqref{eq:T} requires $N\bs_e\sim\text{Multinomial}(N,\bp)$, that is, independence of the bin indicators across seeds. This is a statement about the joint behavior of the realizations relative to the expert's quantiles, and it fails whenever the PITs $U_{ei}=F_{ei}(x_i)$ are dependent across $i$. Dependence arises naturally when the $X_i$ share a common driver. The expert's distributions describe her uncertainty about each $X_i$ marginally, and even an ideal Bayesian expert whose marginals are exactly right has PITs that inherit the residual dependence of the variables given her information (Section~\ref{sec:ideal}). Calibration in the sense of~\eqref{eq:H0} is a marginal property \citep{dawid1984present,gneiting2007probabilistic}; the $\chi^2_{K-1}$ reference adds a joint independence property that calibrated experts need not satisfy.

\section{Methodology: calibration scores under dependent seeds}\label{sec:theory}

\subsection{The limit law of the calibration statistic}\label{sec:limit}

Fix an expert and drop the index $e$. The long-run covariance of the bin-frequency vector is
\begin{equation}\label{eq:SigmaN}
\SigN = \Var\big(\sqrt N\,\bs\big) = \frac1N\sum_{i=1}^N\sum_{j=1}^N\Cov(\bZ_i,\bZ_j),
\end{equation}
and we call $\bm{A}_N=\Dp^{-1/2}\SigN\Dp^{-1/2}$ the generalized design-effect matrix. Since $\one^\top\bZ_i=1$, we have $\SigN\one=\bm 0$, so $\sqrt{\bp}=(\sqrt{p_1},\dots,\sqrt{p_K})^\top$ is an eigenvector of $\bm A_N$ with eigenvalue zero. We denote the remaining eigenvalues by $\lambda_{N,1}\ge\dots\ge\lambda_{N,K-1}\ge 0$. Under independence $\SigN=\Dp-\bp\bp^\top$ and $\bm A_N=\bm I-\sqrt{\bp}\sqrt{\bp}^\top$ is a projection of rank $K-1$, so every $\lambda_{N,k}$ equals one.

Suppose now that $H_0$ holds, that the seed locations form an increasing-domain array with a minimum separation, that the indicator field $\{\bZ_i\}$ satisfies the $\alpha$-mixing conditions of the central limit theorem of \citet[Theorem~1]{jenish2009central}, and that $\SigN$ converges to a matrix $\bm\Sigma$ of rank $K-1$. Then
\begin{equation}\label{eq:limit}
T = 2N\,I(\bs;\bp)\;\xrightarrow{d}\;Q_{\bm\lambda}:=\sum_{k=1}^{K-1}\lambda_k\,\xi_k^2,\qquad \xi_k\overset{\mathrm{iid}}{\sim}\mathcal N(0,1),
\end{equation}
where $\lambda_1\ge\dots\ge\lambda_{K-1}>0$ are the non-zero eigenvalues of $\bm A=\Dp^{-1/2}\bm\Sigma\Dp^{-1/2}$. The derivation (\ref{app:limit}) combines a central limit theorem for $\sqrt N(\bs-\bp)$ with a second-order expansion of the Kullback-Leibler divergence around $\bp$, whose linear term vanishes because $\one^\top(\bs-\bp)=0$. The limit is the multinomial analogue of the Rao-Scott result for survey data \citep{rao1981analysis}, with spatial or temporal mixing in the role of the sampling design, and it reduces to $\chi^2_{K-1}$ exactly when $\bm\Sigma=\Dp-\bp\bp^\top$. The mixing conditions hold, for example, for indicator transforms of Gaussian fields with fast-decaying correlation, because indicators are measurable functions of the field and inherit its mixing coefficients \citep{doukhan1994mixing}.

The direction of the distortion holds for every $N$: whenever $\SigN-(\Dp-\bp\bp^\top)$ is positive semi-definite, which is the case under the non-negative latent correlations of Section~\ref{sec:hermite}, every $\lambda_{N,k}$ is at least one. In the limit $Q_{\bm\lambda}$ then dominates a $\chi^2_{K-1}$ variable pointwise, so that
\begin{equation}\label{eq:anticons}
\lim_{N\to\infty}\Pr\big(C^{\mathrm{cl}}\le\alpha\big)=\Pr\big(Q_{\bm\lambda}\ge\chi^2_{K-1,1-\alpha}\big)>\alpha
\end{equation}
for every $\alpha\in(0,1)$ as soon as one eigenvalue exceeds one (\ref{app:anticons}). Positive dependence among seeds thus makes calibrated experts look miscalibrated, and the classical score overstates the evidence against \emph{every} expert.

\subsection{A Gaussian-copula working model and its Hermite representation}\label{sec:hermite}

To make $\SigN$ computable we model the PITs through a Gaussian copula. Let $W_i=\Phi^{-1}(U_i)$ be the probit of the PIT and $\tau_k=\Phi^{-1}(\pi_k)$ the latent thresholds, with $\tau_0=-\infty$ and $\tau_K=+\infty$, so that $Z_{ik}=\mathbb 1\{\tau_{k-1}<W_i\le\tau_k\}$. The working model is
\begin{equation}\label{eq:copula}
(W_1,\dots,W_N)^\top\sim\mathcal{N}_N(\bm 0,\bm R),\qquad R_{ij}=\rho_{ij}=\rho(\|\bm u_i-\bm u_j\|),
\end{equation}
for a valid correlation function $\rho$. Under~\eqref{eq:copula} the marginal null~\eqref{eq:H0} holds whatever $\bm R$ is, and only bivariate bin probabilities enter $\SigN$, so the model is used solely to describe \emph{pairwise} dependence among bin indicators.

Let $\He_m$ denote the probabilists' Hermite polynomials ($\He_0=1$, $\He_1(x)=x$, $\He_{m+1}(x)=x\He_m(x)-m\He_{m-1}(x)$) and $\varphi$ the standard normal density. Define, for $m\ge1$, the Hermite bin vectors $\bc_m\in\mathbb{R}^K$ and the scalars $g_m$ and $\kappa_{N,m}$ by
\begin{equation}\label{eq:cm}
c_{m,k}=\varphi(\tau_{k-1})\He_{m-1}(\tau_{k-1})-\varphi(\tau_k)\He_{m-1}(\tau_k),\qquad
g_m=\frac{\bc_m^\top\Dp^{-1}\bc_m}{m!},\qquad \kappa_{N,m}=\frac1N\sum_{i\neq j}\rho_{ij}^m ,
\end{equation}
with the convention $\varphi(\pm\infty)\He_{m-1}(\pm\infty)=0$. Mehler's expansion of the bivariate normal density in Hermite polynomials \citep{mehler1866entwicklung,lancaster1958structure} yields, for every pair $i\neq j$ with $|\rho_{ij}|<1$,
\begin{equation}\label{eq:covpair}
\Cov(\bZ_i,\bZ_j)=\sum_{m=1}^\infty\frac{\rho_{ij}^m}{m!}\,\bc_m\bc_m^\top ,
\end{equation}
an absolutely convergent series whose coefficient vectors are the Hermite coefficients of the bin indicators (\ref{app:hermite}). Summing over pairs gives
\begin{equation}\label{eq:SigmaHermite}
\SigN=\Dp-\bp\bp^\top+\sum_{m=1}^\infty\frac{\kappa_{N,m}}{m!}\,\bc_m\bc_m^\top .
\end{equation}
Moreover $\bc_m^\top\one=0$ for every $m$, and by Parseval's identity $\sum_{m\ge1}\bc_m\bc_m^\top/m!=\Dp-\bp\bp^\top$.

Representation~\eqref{eq:SigmaHermite} separates the geometry of the seed set, which enters only through the scalars $\kappa_{N,m}$, from the elicitation format, which enters only through the vectors $\bc_m$ and depends on $\bp$ alone. For symmetric formats, odd-order vectors are antisymmetric and capture common shifts of the realizations relative to the quantiles; even-order vectors are symmetric and capture common dispersion effects, such as realizations being jointly too concentrated in, or too spread out of, the central bins.

\subsection{Design effect and effective number of seeds}\label{sec:deff}

We define the design effect of the seed set as the average non-zero eigenvalue of $\bm A_N$, $\dbar=(K-1)^{-1}\sum_k\lambda_{N,k}=(K-1)^{-1}\tr(\Dp^{-1}\SigN)$, and the effective number of seed variables as $\Neff=N/\dbar$. Taking traces in~\eqref{eq:SigmaHermite} gives the closed form
\begin{equation}\label{eq:deff}
\dbar = 1+\frac{1}{K-1}\sum_{m=1}^\infty g_m\,\kappa_{N,m},\qquad g_m\ge0,\qquad \sum_{m=1}^\infty g_m=K-1 .
\end{equation}
When all latent correlations lie in $[0,1)$, every $\kappa_{N,m}$ is non-negative and $\kappa_{N,m}\le\kappa_{N,1}$, which gives, for every truncation order $M\ge1$,
\begin{equation}\label{eq:bounds}
1+\frac{1}{K-1}\sum_{m=1}^M g_m\kappa_{N,m}\;\le\;\dbar\;\le\;1+\kappa_{N,1},
\end{equation}
together with $\lambda_{N,k}\ge1$ for all $k$ (\ref{app:deff}). The upper bound $1+\kappa_{N,1}=\Var(\sqrt N\,\bar W)$ is Kish's design effect for the mean of the latent probits \citep{kish1965survey}. Coarsening the continuous PITs into $K$ bins therefore attenuates the dependence: the bins never carry more dependence than the latent field, and the Hermite weights of the format determine how much of it survives.

\begin{table}[t]
\centering
\caption{Hermite weights $g_m$ of the 5-50-95 elicitation format ($K=4$, $\bp=(0.05,0.45,0.45,0.05)$) and their cumulative share of the total $K-1=3$. The partial sum over $m\le 2\times10^5$ equals $2.9955$.}\label{tab:gm}
\small
\input{tables/tab_gm.tex}
\end{table}

\begin{figure}[t]
\centering
\includegraphics[width=\textwidth]{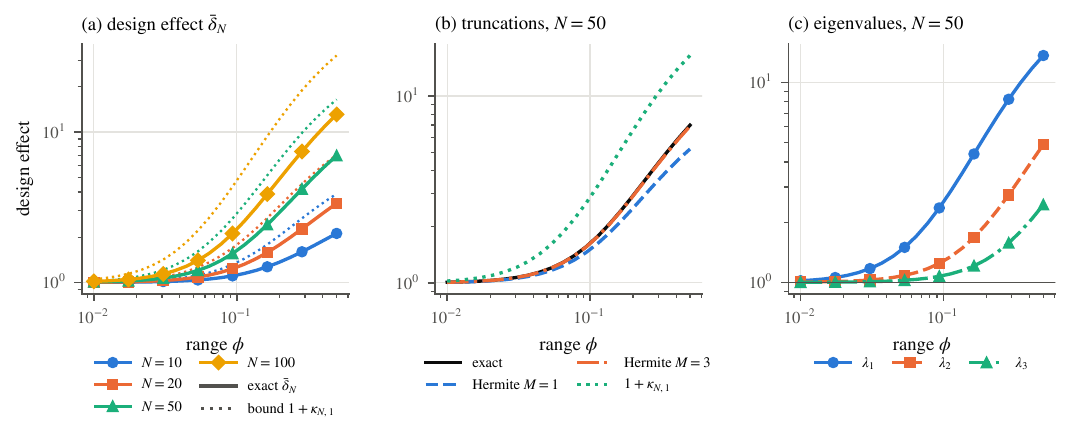}
\caption{Design effect of the seed set under model~\eqref{eq:copula} with $\rho(h)=0.8\exp(-h/\phi)$, for $N$ seeds uniform on the unit square (averages over 30 configurations). (a) $\dbar$ (solid) and the latent upper bound $1+\kappa_{N,1}$ (dotted). (b) Exact $\dbar$ against Hermite truncations and bounds, $N=50$. (c) Eigenvalues $\lambda_{N,k}$ of the generalized design-effect matrix, $N=50$.}\label{fig:deff}
\end{figure}

Table~\ref{tab:gm} reports $g_1,\dots,g_8$ for Cooke's format. The first-order term accounts for 27.1\% of the total weight and the first three terms for 63.5\%. The fourth weight is almost zero ($g_4=0.0046$) because every threshold lies at or near a root of $\He_3(x)=x^3-3x$: the median threshold $\tau_2=0$ is a root, and $\tau_1,\tau_3=\mp1.645$ are close to the roots $\mp\sqrt3\approx\mp1.732$, so that $\varphi(\tau_k)\He_3(\tau_k)$ is small at every threshold and $|c_{4,k}|=0.050$ for every $k$. The total is reached slowly, because the Hermite coefficients of an indicator decay slowly; but $\kappa_{N,m}$ also decays geometrically in $m$, and the truncation with $M=3$ reproduces $\dbar$ to within 1\% over the range of dependence considered in Section~\ref{sec:sim} (Figure~\ref{fig:deff}b).

Figure~\ref{fig:deff} shows how fast the design effect grows. With $N=50$ seeds and range $\phi=0.1$, so that correlations fall to about $0.8e^{-1}\approx0.3$ at one tenth of the side of the domain, $\dbar\approx1.6$: thirty effective seeds out of fifty. With $\phi=0.3$ the design effect reaches about 4.4. The latent bound $1+\kappa_{N,1}$ overstates $\dbar$ by a factor between 1.6 and 2.4 once $\phi\ge0.08$, and the three eigenvalues are markedly unequal, which is why first-order corrections are not sufficient on their own (Section~\ref{sec:corrected}).

The lower bound in~\eqref{eq:bounds} with $M=1$ also implies a ceiling under infill sampling, in which seeds are added within a fixed region. If every latent correlation is at least $\rho_{\min}>0$, then $\dbar\ge1+g_1(N-1)\rho_{\min}/(K-1)$, and since $g_1<K-1$,
\begin{equation}\label{eq:ceiling}
\Neff<\frac{K-1}{g_1\,\rho_{\min}}\qquad\text{for every }N,
\end{equation}
which equals $3.684/\rho_{\min}$ for Cooke's format (\ref{app:ceiling}). Adding seeds from the same field does not buy unlimited discriminating power: with a minimum residual correlation of $0.25$, no number of seeds yields more than about 15 effective seeds. In this regime $\bs$ does not even converge to $\bp$ and asymptotic $\chi^2$ arguments fail, whereas the parametric bootstrap of Section~\ref{sec:corrected} remains valid under the working model.

\subsection{Ideal experts have dependent PITs}\label{sec:ideal}

Model~\eqref{eq:copula} is not an ad hoc device. Let $X_i=\mu(\bm u_i)+S(\bm u_i)+\varepsilon_i$, with $S$ a stationary Gaussian field and $\varepsilon_i$ white noise. Consider an expert who knows $\mu$ and the covariance structure and has observed data $\mathcal{Y}_e$, such as noisy measurements of $X$ at some sites. If she reports the exact posterior predictive marginals of $X_i\mid\mathcal Y_e$, she is perfectly calibrated in the sense of~\eqref{eq:H0}. Her PIT probits $W_i=\{X_i-\E(X_i\mid\mathcal Y_e)\}/\mathrm{sd}(X_i\mid\mathcal Y_e)$ are nevertheless jointly Gaussian with the correlation matrix of the posterior, which is not diagonal unless $\mathcal{Y}_e$ determines $S$ at every seed. Model~\eqref{eq:copula} is thus exact for this ideal expert, and better-informed experts have weaker residual dependence and a smaller design effect.

\subsection{Corrected calibration scores}\label{sec:corrected}

Given the eigenvalues $\bm\lambda=(\lambda_1,\dots,\lambda_{K-1})$, or an estimate of them, we consider four corrected scores.

\paragraph{First-order Rao-Scott, or effective-seed, score}
\begin{equation}\label{eq:rs1}
C^{\mathrm{RS1}}=1-F_{\chi^2_{K-1}}\!\big(T/\dbar\big)=1-F_{\chi^2_{K-1}}\!\big(2\Neff\,I(\bs;\bp)\big).
\end{equation}
The second expression shows that $C^{\mathrm{RS1}}$ is \emph{exactly} the classical score computed with the number of seeds replaced by $\Neff$. This is the ``calibration power'' option of the Excalibur software with power $\beta=1/\dbar$, where the power ``adjusts the effective number of seed variables, as a percentage of the total number'' \citep{cooke2000procedures}; the open-source Anduryl implementation computes the statistic as $2N\beta I(\bs;\bp)$ \citep{rongen2020update}, building on ANDURIL \citep{leontaris2018anduril}. Our analysis gives this long-standing device a precise statistical meaning and a way to estimate it. The score matches the first moment of the limit~\eqref{eq:limit}, since $\E\,Q_{\bm\lambda}=(K-1)\dbar$.

\paragraph{Second-order (Satterthwaite) score}
Matching the first two moments of $Q_{\bm\lambda}$ by $\gamma\chi^2_\nu$ \citep{satterthwaite1946approximate,rao1984chi} gives $\gamma=\dbar(1+a^2)$ and
\begin{equation}\label{eq:rs2}
C^{\mathrm{RS2}}=1-F_{\chi^2_\nu}\!\Big(\frac{T}{\dbar(1+a^2)}\Big),\quad
a^2=\frac{\sum_k(\lambda_k-\dbar)^2}{(K-1)\dbar^2},\quad \nu=\frac{K-1}{1+a^2},
\end{equation}
where $a$ is the coefficient of variation of the eigenvalues.

\paragraph{Exact-limit score}
$C^{\mathrm{IM}}=\Pr(Q_{\bm\lambda}>T)$, computed by Imhof's inversion formula \citep{imhof1961computing} or approximated as in \citet{liu2009new}.

\paragraph{Parametric-bootstrap score}
Simulate $\bm W^{(b)}\sim\mathcal N_N(\bm 0,\bm R)$, bin it at the null thresholds $\tau_k$, compute $T^{(b)}$ for $b=1,\dots,B$, and set
\begin{equation}\label{eq:pb}
C^{\mathrm{PB}}=\frac{1+\#\{b:T^{(b)}\ge T\}}{B+1}.
\end{equation}
With $\bm R$ known, $C^{\mathrm{PB}}$ is a valid $p$-value in finite samples under the working model. It also accounts for the discreteness of $T$ at small $N$, which distorts even the classical score under independence, and it remains valid under infill. With $\bm R$ replaced by an estimate, its validity becomes asymptotic, as for the plug-in scores below.

When $\hat{\bm\lambda}$ is consistent for $\bm\lambda$, which with a single set of realizations requires increasing-domain asymptotics under which the dependence parameters are consistently estimated, the exact-limit score $C^{\mathrm{IM}}(\hat{\bm\lambda})$ is asymptotically uniform under $H_0$, and $C^{\mathrm{RS1}}$ and $C^{\mathrm{RS2}}$ have the same limits as the corresponding scores computed with $\bm\lambda$; $C^{\mathrm{RS1}}$ is asymptotically uniform only if all eigenvalues are equal (\ref{app:valid}). For a fixed $\bm\lambda$, $C^{\mathrm{RS1}}$, $C^{\mathrm{RS2}}$ and $C^{\mathrm{IM}}$ are strictly decreasing in $T$, so they rank experts with the same design effect exactly as the classical score does. What changes is their numerical value, which is what enters the weights~\eqref{eq:weights} and the cut-off optimization, and the ranking between experts whose residual dependence differs, such as better- and less-informed experts (Section~\ref{sec:ideal}).
\section{Estimation of the dependence and computation}\label{sec:estimation}

\subsection{The identifiability problem}

The dependence parameters must be estimated from the same seeds that are used for scoring, and the fact that all experts face a single set of realizations creates a genuine identifiability problem. A common upward shift of all realizations relative to an expert's medians may be a bias of the expert, which violates $H_0^{(e)}$; it may equally be a common positive deviation of the seed variables from what was expected, which is part of the null variability. With one set of realizations, the two explanations cannot be fully separated. An estimator of dependence that used the null margins, that is, the fixed thresholds $\tau_k$, would interpret any bias as dependence, inflating $\dbar$ precisely for the experts who are biased and partially excusing them. We therefore require the dependence estimate to be invariant to location-scale miscalibration.

\subsection{REML on rank normal scores}

Let $\tilde U_{ei}$ be the PIT of $x_i$ under expert $e$'s minimum-information distribution on $[L_i,U_i]$, which is computable from the elicited quantiles and lies in $(0,1)$. Let $r_{ei}$ be its rank among $\tilde U_{e1},\dots,\tilde U_{eN}$, and define the rank normal scores $\tilde W_{ei}=\Phi^{-1}\{r_{ei}/(N+1)\}$, rescaled to unit sample variance. Ranks are invariant to any strictly increasing transformation common to an expert's seeds, so $\tilde{\bm W}_e$ is unchanged if the expert's latent probits are replaced by $(W_{ei}-b_e)/\sigma_e$ for any bias $b_e$ and any over- or under-confidence factor $\sigma_e>0$ (\ref{app:inv}). This mirrors the semi-parametric copula estimators of \citet{genest1995semiparametric}.

We parametrize the correlation matrix of expert $e$ as $\bm R_e=\bm R(\phi,\nu_e)$, with entries $\rho_{e,ij}=\nu_e\exp(-h_{ij}/\phi)$ for $i\neq j$ and $\rho_{e,ii}=1$, where $h_{ij}$ is the distance between seeds $i$ and $j$. Distinct seeds at distance zero, which occur in the question spaces of Section~\ref{sec:qspace}, therefore have correlation $\nu_e$. The common range $\phi$ reflects the shared structure of the seed variables, and the expert-specific share $\nu_e\in[0,1)$ reflects the fact that better-informed experts have weaker residual dependence (Section~\ref{sec:ideal}). Normal scores are centered and scaled by construction, which removes the common deviation of the realizations over the seed set; this is the component that drives the calibration statistic, so plain maximum likelihood underestimates the dependence, much as it underestimates variance components when a mean is estimated \citep{patterson1971recovery,harville1977maximum}. We therefore use the restricted likelihood with an expert-specific unknown mean and variance,
\begin{equation}\label{eq:reml}
\ell_R(\phi,\bm\nu)=\sum_{e=1}^E\Big[-\tfrac12\log|\bm R_e|-\tfrac12\log\big(\one^\top\bm R_e^{-1}\one\big)-\tfrac{N-1}{2}\log\big(\tilde{\bm W}_e^\top\bm P_e\tilde{\bm W}_e\big)\Big],
\end{equation}
where $\bm P_e=\bm R_e^{-1}-\bm R_e^{-1}\one\one^\top\bm R_e^{-1}/(\one^\top\bm R_e^{-1}\one)$, with the expert-specific variance profiled out analytically. Summing over experts is a composite likelihood, because experts facing the same realizations are not independent of one another; this affects the efficiency of the estimator, not the unbiasedness of its estimating equations \citep{varin2011overview}. For fixed $\phi$, \eqref{eq:reml} separates across experts, so each $\nu_e$ is maximized by a one-dimensional search and $\phi$ by a profile search (\ref{app:comp}). We constrain the practical range $3\phi$ not to exceed the diameter of the seed set. Longer ranges cannot be identified from one set of realizations, since within the seed set the dependence is then indistinguishable from a constant shift, that is, from a common bias of the expert. The constraint removes a heavy right tail of the estimated design effect and leaves its median essentially unchanged. In computation $\nu_e$ is also restricted to $[0,0.98]$, to keep $\bm R_e$ well conditioned. Both constraints can bind in applications, and Section~\ref{sec:app_cases} reports the sensitivity of the results to them. The plug-in matrices $\hat{\bm R}_e=\bm R(\hat\phi,\hat\nu_e)$ then feed~\eqref{eq:SigmaN}, evaluated at the null thresholds, and the scores of Section~\ref{sec:corrected}.

When only bin labels are available, as in some published data sets, a two-stage pairwise composite likelihood can be used instead \citep{heagerty1998composite}: thresholds are first estimated freely from each expert's marginal bin frequencies, which preserves invariance to miscalibration, and the bivariate rectangle probabilities of~\eqref{eq:copula} are then maximized over nearest-neighbor pairs. This estimator is implemented in Supplementary material S1; it is less efficient, because the bins discard the ordering of realizations within a bin.

\subsection{Seeds without coordinates: a question space}\label{sec:qspace}

Seed variables rarely come with geographical coordinates, but they usually come with \emph{attributes}: the quantity measured, the region, the period, the population. We place seed $i$ at the attribute vector $\bm a_i=(a_{i1},\dots,a_{iJ})$ and use the Hamming distance $h_{ij}=\sum_{l=1}^J\mathbb 1\{a_{il}\neq a_{jl}\}$, so that $\rho_{ij}=\nu\exp(-h_{ij}/\phi)$ decays with the number of attributes in which two seeds differ. Seeds sharing every attribute but one, such as the same indicator in two regions in the same period, are the most strongly dependent. A single attribute ($J=1$) gives an exchangeable within-block structure as a special case. The matrix with entries $\exp(-h_{ij}/\phi)$ is positive semi-definite: it is the elementwise product, over attributes, of the matrices with entries $\exp(-\mathbb 1\{a_{il}\neq a_{jl}\}/\phi)$, each of which is a convex combination of the all-ones matrix and a block-diagonal matrix of ones, and the Schur product theorem applies. With $\nu_e<1$ the matrix $\bm R_e$ is therefore positive definite. Because the choice of attributes is a modeling decision, we report results for the full attribute set and for each attribute alone (Section~\ref{sec:app}). When the attributes of the seeds are not recorded, a question space can also be derived from latent structures among the items, such as the communities recovered by network analyses of Delphi panels \citep{calleo2026hybrid,calleo2026community}. The restricted likelihood also provides a check of the need for a correction: the statistic $\Lambda=2\{\ell_R(\hat\phi,\hat{\bm\nu})-\ell_R(\hat\phi,\bm 0)\}$ compares the fitted dependence with independence, and we calibrate it by a parametric bootstrap under independence, since $\bm\nu=\bm 0$ lies on the boundary of the parameter space. All question spaces share the parameters $(\phi,\bm\nu)$ and the restricted likelihood under independence, so their values of $\Lambda$ can be compared directly.

\subsection{Propagating the estimation uncertainty}\label{sec:prepivot}

The plug-in score $C^{\mathrm{PB}}(\hat{\bm R})$ ignores the sampling variability of $(\hat\phi,\hat{\bm\nu})$, which is substantial when all experts face the same realizations. We propagate it by prepivoting \citep{beran1988prepivoting}. Let $\hat p_e=C^{\mathrm{PB}}_e(\hat{\bm R}_e)$. We simulate $R$ null panels from the fitted model, each with a common latent component of correlation $\exp(-h/\hat\phi)$ loaded with weight $\sqrt{\hat\nu_e}$ and independent expert-specific noise. On each simulated panel we re-estimate the dependence and recompute the bootstrap scores $p^\ast_{e,r}$. The calibrated score is
\begin{equation}\label{eq:pbcal}
C^{\mathrm{PBc}}_e=\frac{1+\#\{(e',r):p^\ast_{e',r}\le\hat p_e\}}{1+ER}.
\end{equation}
If the plug-in score were exactly uniform under the fitted model, $C^{\mathrm{PBc}}$ would coincide with it. When the estimator underestimates the dependence, the simulated $p^\ast$ are too often small, and~\eqref{eq:pbcal} raises the score accordingly. The cost is $R+1$ model fits per panel; we use $R=49$ and $B=199$.

\begin{algorithm}[t]
\caption{Dependence-robust calibration scores}\label{alg:drcm}
\begin{algorithmic}[1]
\Require quantiles $q_{eik}$, realizations $x_i$, seed coordinates or attributes $\bm u_i$, levels $\pi_k$
\State Compute bins $\bZ_{ei}$, frequencies $\bs_e$ and $T_e=2NI(\bs_e;\bp)$ for every expert
\State Compute minimum-information PITs $\tilde U_{ei}$ and rank normal scores $\tilde{\bm W}_e$
\State Maximize the restricted likelihood~\eqref{eq:reml} over $(\phi,\nu_1,\dots,\nu_E)$, with $3\phi$ at most the diameter of the seed set
\For{$e=1,\dots,E$}
  \State $\hat{\bm R}_e\gets\bm R(\hat\phi,\hat\nu_e)$; $\hat{\bm\Sigma}_{N,e}\gets$~\eqref{eq:SigmaN} under~\eqref{eq:copula} at thresholds $\tau_k=\Phi^{-1}(\pi_k)$
  \State $\hat{\bm\lambda}_e\gets$ non-zero eigenvalues of $\Dp^{-1/2}\hat{\bm\Sigma}_{N,e}\Dp^{-1/2}$; $\hat{\bar\delta}_e\gets\mathrm{mean}(\hat{\bm\lambda}_e)$; $\hat N_{\mathrm{eff},e}\gets N/\hat{\bar\delta}_e$
  \State Report $C_e^{\mathrm{RS2}}$ from $\hat{\bm\lambda}_e$ and $C_e^{\mathrm{PB}}$ from $B$ draws of $\mathcal N_N(\bm 0,\hat{\bm R}_e)$
\EndFor
\State Optionally calibrate the scores by prepivoting,~\eqref{eq:pbcal}
\State Use the corrected scores in~\eqref{eq:weights}, and score the DM with $\bm R(\hat\phi,\bar{\hat\nu})$
\end{algorithmic}
\end{algorithm}
\section{Simulation study}\label{sec:sim}

\subsection{Design}\label{sec:design}

Seed locations are drawn uniformly on the unit square. The latent probits of the experts' PITs are
\begin{equation}\label{eq:dgp}
W_{ei}=\sqrt{\nu}\,S(\bm u_i)+\sqrt{1-\nu}\,\varepsilon_{ei},
\end{equation}
where $S$ is a zero-mean, unit-variance Gaussian field with correlation $\exp(-h/\phi)$ and $\varepsilon_{ei}$ is i.i.d.\ $\mathcal N(0,1)$. The field $S$ is \emph{shared} by all experts, as it is when the seeds have a common driver, while $\varepsilon_{ei}$ is expert-specific. We take $\nu=0.8$ and $\phi\in\{0.05,0.1,0.2,0.3\}$, so that the practical range $3\phi$ runs from 15\% to 90\% of the side of the domain, plus a scenario $\phi=0$ with fully independent seeds. We use $N\in\{10,20,30,50,100\}$ seeds and panels of $E=6$ experts. An expert with bias $b$ and scale factor $\sigma$ has latent probit $(W_{ei}-b)/\sigma$ relative to her own quantiles, so that $b\neq0$ means systematically shifted medians and $\sigma<1$ means overconfidence; calibrated experts have $(b,\sigma)=(0,1)$.

For each panel we compute the classical score~\eqref{eq:T}; the corrected scores RS1, RS2 and PB with dependence estimated by REML on rank normal scores (Section~\ref{sec:estimation}); and two \emph{oracle} versions, RS1$^\ast$ and PB$^\ast$, that use the true correlation matrix. Parametric bootstraps use $B=499$ draws. The size study uses 600 panels per cell (300 for $N=100$), that is, 3600 (1800) expert-level scores. The Monte Carlo standard error of a rejection rate near 5\% is about 0.4 (0.5) percentage points, ignoring the within-panel correlation induced by the shared field, which inflates it somewhat.

\subsection{Size: how often are calibrated experts rejected?}\label{sec:size}

\begin{figure}[t]
\centering
\includegraphics[width=\textwidth]{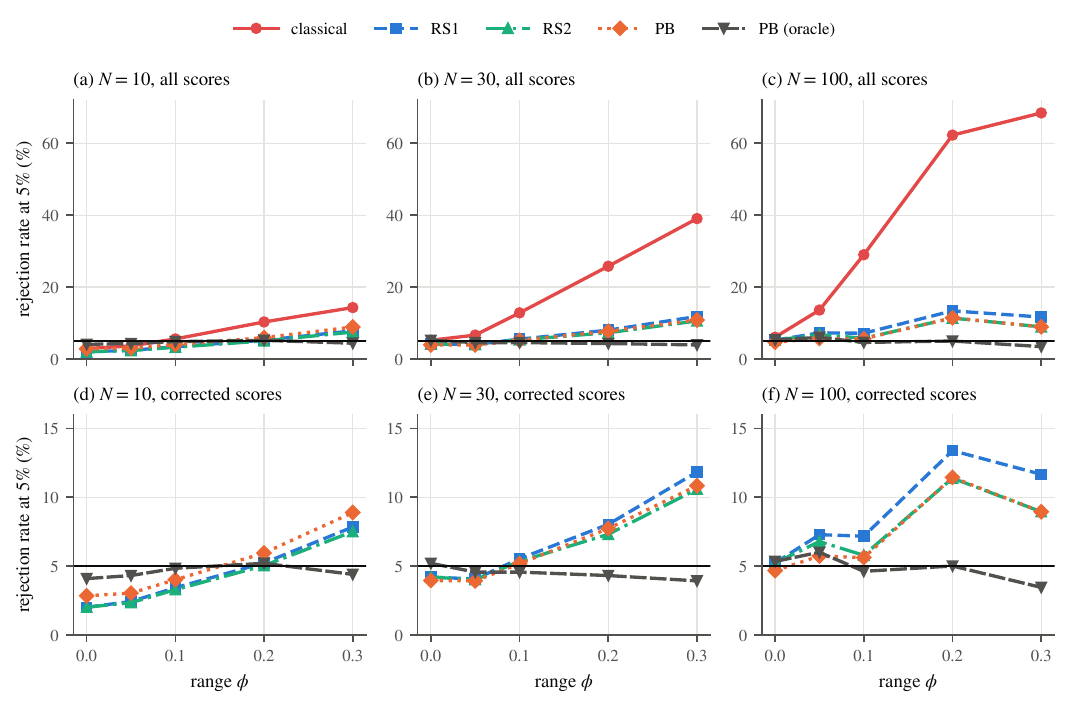}
\caption{Rejection rate at the 5\% level of calibrated experts as a function of the range $\phi$, for $N=10,30,100$ seeds. Top row (a-c): all scores. Bottom row (d-f): corrected scores only, on an expanded vertical scale. The horizontal line marks the nominal 5\%. ``PB (oracle)'' uses the true dependence; the other corrected scores use the REML estimate.}\label{fig:size}
\end{figure}

Table~\ref{tab:size} and Figure~\ref{fig:size} report the percentage of calibrated experts whose score falls below 0.05, that is, the probability that a calibrated expert is ``rejected'' at the conventional 5\% level. With independent seeds the classical score behaves as expected. It is somewhat conservative for $N\le20$ because of the discreteness of $T$ (2.9\% at $N=10$) and close to nominal for larger $N$ (5.2-6.1\%). As soon as the seeds are dependent it becomes markedly anti-conservative, and the distortion grows with \emph{both} the range and the number of seeds. At $\phi=0.2$ the rejection rate is 10.3\% with 10 seeds, 25.8\% with 30, 40.3\% with 50 and 62.3\% with 100; at $\phi=0.3$ it reaches 68.4\%. At the 1\% level the classical score rejects up to 55.7\% of calibrated experts (\ref{app:extra}, Table~\ref{tab:size01}). The growth with $N$ illustrates~\eqref{eq:deff} and~\eqref{eq:ceiling}: adding seeds with a common driver adds apparent evidence much faster than effective evidence, and with $N=100$ and $\phi=0.3$ the effective number of seeds is only $100/7.86\approx13$.

\begin{table}[t]
\centering
\caption{Size study: percentage of calibrated experts with calibration score below 0.05 (nominal level 5\%). $\bar\delta_N$ is the true design effect averaged over seed configurations and $\widehat{\bar\delta}_N$ the median REML estimate. RS1$^\ast$ and PB$^\ast$ use the true correlation; RS1, RS2 and PB use the estimated one. Classical rates above 7.5\% are in bold. Each cell is based on 3600 expert-level scores (1800 for $N=100$).}\label{tab:size}
\footnotesize
\input{tables/tab_size.tex}
\end{table}

The oracle corrections confirm the theory. The bootstrap score with known dependence stays between 3.4\% and 6.0\% in all 25 settings. The first-order correction RS1$^\ast$ is accurate except when the eigenvalues are very unequal: at $N=100$ and $\phi=0.2$ it rejects 7.7\% of calibrated experts, the expected effect of matching only the first moment. With \emph{estimated} dependence the corrected scores are close to nominal for weak to moderate dependence: for $\phi\le0.1$ the bootstrap score ranges from 2.8\% to 6.0\% in every setting. For stronger dependence, a residual excess remains, 5.9-12.9\% at $\phi\in\{0.2,0.3\}$, which stems from estimation uncertainty and from a moderate downward bias of $\widehat{\bar\delta}_N$ in the most dependent settings (for example, a median of 3.79 against a true 4.40 at $N=50$, $\phi=0.3$). In these settings the corrected scores still remove between 58\% and 94\% of the classical score's excess rejection, and more than 80\% in eight of the ten cells. RS2 is never worse than RS1, and PB and RS2 are within about one percentage point of each other except at $N=10$; both clearly improve on RS1 for $N\ge50$.

\begin{figure}[t]
\centering
\includegraphics[width=\textwidth]{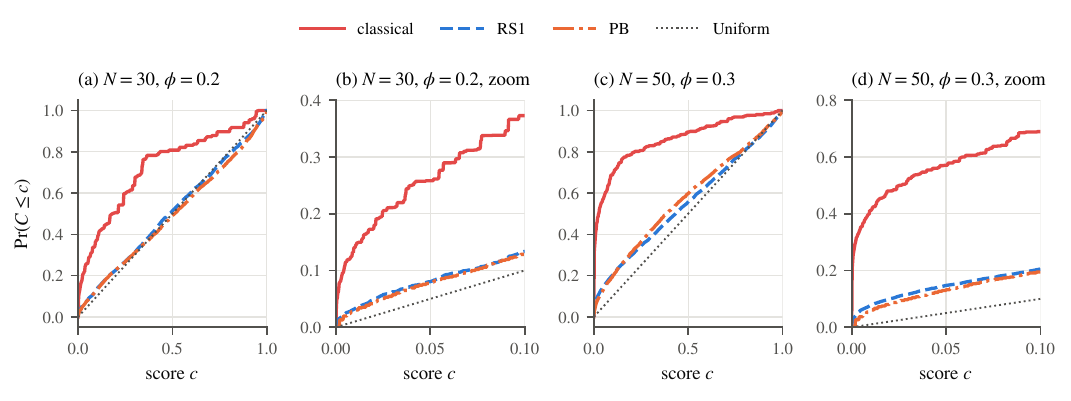}
\caption{Empirical distribution functions of the calibration scores of calibrated experts, for $(N,\phi)=(30,0.2)$ in (a, b) and $(50,0.3)$ in (c, d). A valid score follows the diagonal (Uniform). Panels (b) and (d) zoom on $c\le0.1$.}\label{fig:ecdf}
\end{figure}

Figure~\ref{fig:ecdf} shows the whole distribution of the scores of calibrated experts, which should be uniform. With $N=50$ and $\phi=0.3$, half of all calibrated experts receive a classical score below 0.026, against a median of 0.5 for a valid score; the median bootstrap score is 0.38. The classical scores of genuinely calibrated experts are thus compressed towards zero, which matters for weighting even when no formal test is performed, because the weights~\eqref{eq:weights} are proportional to the scores.

\subsection{Size after prepivoting}\label{sec:sim_prepivot}

Table~\ref{tab:prepivot} compares the plug-in bootstrap score with its prepivoted version~\eqref{eq:pbcal} on 200 panels per setting (1200 expert-level scores, Monte Carlo standard error about 0.6 percentage points near 5\%). Both use $B=199$ draws, and prepivoting uses $R=49$ simulated panels, so the plug-in rates differ slightly from those of Table~\ref{tab:size}, which used $B=499$ and more panels. Prepivoting reduces the residual excess rejection where it matters, under strong dependence: from 10.1\% to 8.3\% at $\phi=0.2$, from 10.1\% to 8.6\% at $\phi=0.3$, and from 14.1\% to 10.5\% with $N=50$ and $\phi=0.3$. Under weak or no dependence it changes little (3.2\% and 6.1\% against 2.8\% and 5.2\%), within Monte Carlo error. The gain is real but partial, because the simulated panels inherit the single-realization limitation of the observed one. Prepivoting is therefore worth its cost of about fifty model fits per panel when the estimated design effect is large, and unnecessary otherwise.

\begin{table}[t]
\centering
\caption{Plug-in (PB) and prepivoted (PBc) bootstrap scores: percentage of calibrated experts with score below 0.05 ($E=6$, 200 panels per setting, $B=199$, $R=49$).}\label{tab:prepivot}
\footnotesize
\input{tables/tab_prepivot.tex}
\end{table}

\subsection{Robustness to the working model}\label{sec:robust}

The corrections rest on the Gaussian copula~\eqref{eq:copula}. To probe this assumption we generate the PITs from two non-Gaussian dependence structures, keeping the Gaussian panel~\eqref{eq:dgp} as a latent layer $V_{ei}$, with $N=30$ and 400 panels per cell. In the first, $W_{ei}=\Phi^{-1}\{F(V_{ei}+1)^2\}$, where $F$ is the non-central $\chi^2_1$ distribution function with non-centrality one; this yields asymmetric dependence. In the second, $W_{ei}=\Phi^{-1}\{F_{t_4}(V_{ei}/\sqrt{G/4})\}$ with $G\sim\chi^2_4$ common to the whole panel, a Student-$t$ copula with tail dependence in which every realization is scaled by the same random factor. The scores are computed exactly as before, that is, under a misspecified Gaussian working model.

Table~\ref{tab:robust} shows that the correction is robust to the asymmetric $\chi^2$ dependence: the rejection rates of RS2 and PB (4.3-12.9\%) are close to those obtained under the Gaussian model, while the classical score rejects up to 31.8\%. The common-scale $t$ copula is a harder case. The corrected scores reduce the rejection rate from 28.5-51.7\% to 14.4-20.1\%, but do not restore it. This is not a defect of the estimator so much as a limit of identifiability. A random scale factor shared by all seeds makes the realizations jointly too dispersed or too concentrated relative to every expert's quantiles. For a single set of realizations this is observationally equivalent to common over- or under-confidence, and the rank normal scores are by design blind to it. Dependence that acts through a common \emph{scale} rather than a common \emph{location} can be corrected only with external information, for example from historical replicates of the seed variables.

\begin{table}[t]
\centering
\caption{Robustness to the dependence model ($N=30$, $E=6$; 400 panels per cell for the non-Gaussian copulas). Percentage of calibrated experts with score below 0.05. The corrected scores use the Gaussian working model with dependence estimated by REML.}\label{tab:robust}
\footnotesize
\input{tables/tab_robust.tex}
\end{table}

\subsection{Power: are miscalibrated experts still detected?}\label{sec:power}

\begin{figure}[t]
\centering
\includegraphics[width=\textwidth]{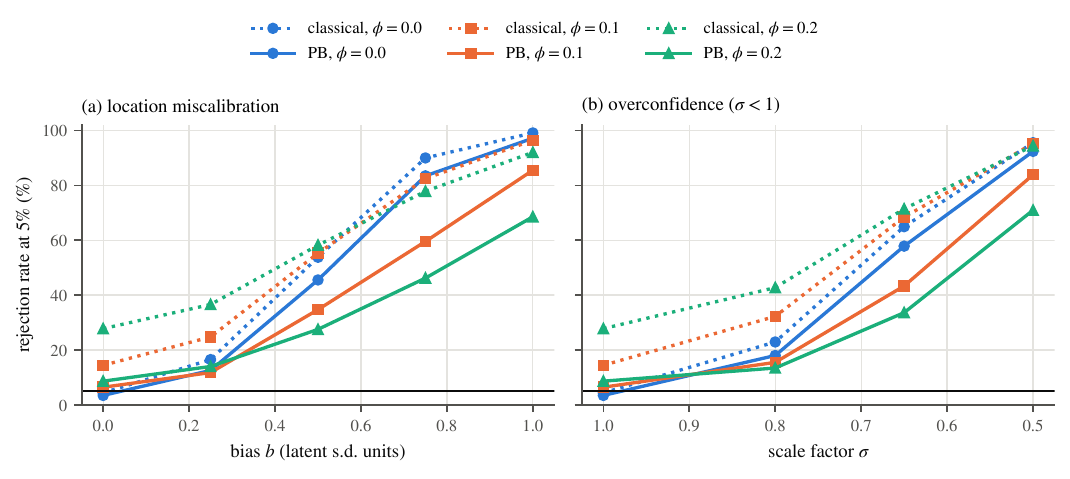}
\caption{Rejection rate at the 5\% level as a function of (a) bias and (b) overconfidence, for $N=30$ seeds. Dotted lines: classical score; solid lines: bootstrap score with estimated dependence. Colors and markers identify the range $\phi$.}\label{fig:power}
\end{figure}

Table~\ref{tab:power} and Figure~\ref{fig:power} report rejection rates at the 5\% level for panels of experts with bias $b$ or scale factor $\sigma$, with $N=30$ seeds and 500 panels per cell. Two facts organize the results.

First, under dependence the classical score's apparent power is partly an artifact of its size distortion: at $\phi=0.2$ it rejects 27.8\% of calibrated experts. For a fixed dependence structure, every score is a monotone function of $T$, so the \emph{size-adjusted} power of the classical score equals that of the oracle score, and any rejections beyond it are false rejections. The oracle column RS1$^\ast$ therefore shows how much information about calibration the seeds really contain. For a bias of half a latent standard deviation, the oracle power drops from 53.8\% with independent seeds to 39.5\% at $\phi=0.1$ and 26.0\% at $\phi=0.2$. For an overconfident expert whose 90\% intervals have 65\% of the correct width ($\sigma=0.65$) it drops from 65.0\% to 51.0\% and 33.7\%. This loss is the honest consequence of having fewer effective seeds.

Second, estimating the dependence costs little power once dependence is present. At $\phi=0.1$ and $\phi=0.2$ the bootstrap score with estimated dependence is within about ten percentage points of the oracle in every cell, and often much closer (for example, 27.5\% against 26.0\% for $b=0.5$, and 71.0\% against 73.2\% for $\sigma=0.5$, at $\phi=0.2$). Estimating a dependence that does not exist has a moderate price: with independent seeds the bootstrap score detects a bias of $0.5$ in 45.5\% of cases against 53.8\% for the classical score, and it is slightly conservative under the null (3.4\% against 4.3\%), because spurious positive estimates of $\nu_e$ push $\widehat{\bar\delta}_N$ above one. When there is external evidence that the seeds are independent, the classical score remains the right choice.

\begin{table}[t]
\centering
\caption{Power study ($N=30$, $E=6$, 500 panels per cell): percentage of experts with calibration score below 0.05. Rows with $b=0$, $\sigma=1$ are calibrated experts (size). Cl.: classical; RS1$^\ast$: oracle first-order correction (under independence it coincides with Cl.); RS1 and PB use the estimated dependence.}\label{tab:power}
\footnotesize
\input{tables/tab_power.tex}
\end{table}

\subsection{Consequences for performance weights and the decision maker}\label{sec:weights}

The last study embeds the scores in the full Classical Model, with $E=8$ experts and $N=20$ seeds whose realizations are $X_i=\sqrt\nu S(\bm u_i)+\sqrt{1-\nu}\,\varepsilon_i$ ($\nu=0.8$). We consider two panels. In panel A, half of the experts are miscalibrated. E1-E2 are calibrated \emph{generalists}, who report $\mathcal N(0,1)$ for every seed. E3-E4 are calibrated \emph{informed} experts, each of whom observes $X_i+\eta_{ei}$ with $\eta_{ei}\sim\mathcal N(0,0.5)$ and reports the exact conditional distribution, so that by Section~\ref{sec:ideal} their residual dependence is weaker. E5 is biased ($b=0.7$), E6 is overconfident ($\sigma=0.5$), E7 is informed but overconfident (posterior standard deviation halved), and E8 is both biased and overconfident ($b=-0.5$, $\sigma=0.6$). In panel B, which is mostly calibrated, E1-E2 are generalists, E3-E6 are informed experts with noise variances $0.3,0.5,0.8,1.2$, E7 is mildly biased ($b=0.4$) and E8 mildly overconfident ($\sigma=0.75$). Information scores, intrinsic ranges and the optimized DM follow Section~\ref{sec:cm}, and for the corrected schemes the DM is scored with the corrected score of the same type. We also report un-optimized global weights ($\alpha=0$). Out-of-sample performance is measured on 40 target variables from an independent realization of the field, which represents the generalization to new questions that performance weighting is meant to deliver. We use the mean quantile (pinball) score at the three levels, a proper scoring rule for quantiles \citep{gneiting2007strictly}, relative to equal weights. There are 400 replicates per range.

\begin{table}[t]
\centering
\caption{Performance weighting ($E=8$, $N=20$, 400 replicates per range). $w_{\mathrm{mis}}$: mean total weight on the miscalibrated experts (four in panel A, two in panel B); $E_{\mathrm{eff}}=1/\sum_ew_e^2$: mean effective number of experts; Rel.\ QS: mean out-of-sample quantile score on new questions relative to equal weights (values below one favor the scheme).}\label{tab:weights}
\footnotesize
\input{tables/tab_weights.tex}
\end{table}

The corrected scores change who is penalized. At $\phi=0.3$ the classical score assigns a value below 0.05 to 25.5\% of the calibrated generalists in both panels, against 9.1-9.2\% for the bootstrap score. The estimated design effects reproduce the ordering predicted in Section~\ref{sec:ideal}, with 1.98 for generalists against 1.45 for informed experts in panel A at $\phi=0.3$. The classical score therefore not only penalizes calibrated experts but penalizes uninformed calibrated experts most, because their PITs inherit more of the shared dependence. At the same time, the corrected score detects miscalibrated experts less often (55.9\% against 76.1\% in panel A, and 21.0\% against 41.6\% in panel B, at $\phi=0.3$), in line with the smaller effective number of seeds.

\begin{figure}[t]
\centering
\includegraphics[width=\textwidth]{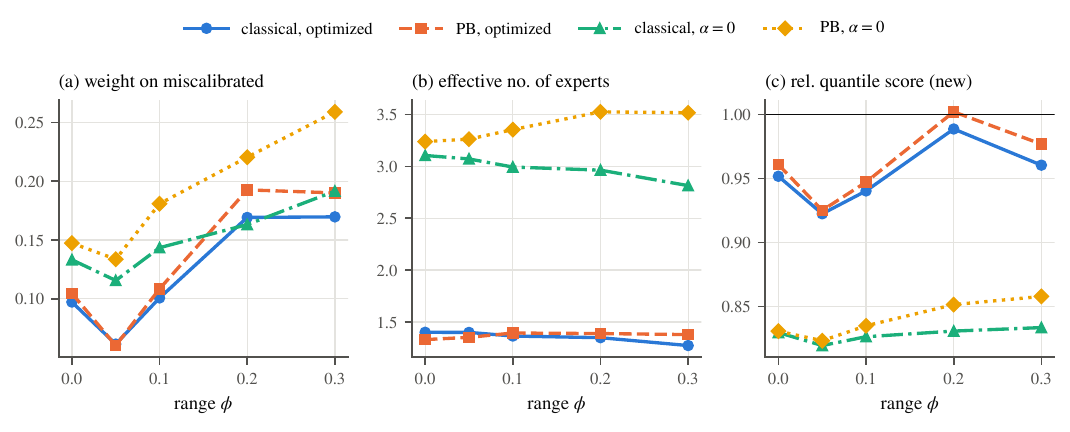}
\caption{Performance weighting in panel A as a function of the range $\phi$: (a) total weight on the miscalibrated experts; (b) effective number of experts; (c) out-of-sample quantile score on new questions relative to equal weights (below one is better).}\label{fig:weights}
\end{figure}

The effect on the DM is more nuanced (Table~\ref{tab:weights} and Figure~\ref{fig:weights}). With optimized weights, the Classical Model concentrates almost all weight on one or two experts under every scheme ($E_{\mathrm{eff}}\approx1.2$-$1.5$). With un-optimized weights the corrected schemes spread the weight more evenly and give more weight to the miscalibrated experts, as the weaker evidence against them warrants. Out of sample, the un-optimized DMs clearly beat equal weights in both panels (relative quantile scores 0.81-0.87), whereas the optimized DMs gain little or nothing. In both panels, the corrected schemes perform similarly to, or slightly worse than, the classical one: for example, for $\alpha=0$ at $\phi=0.3$ the relative score is 0.858 (PB) against 0.834 (classical) in panel A and 0.829 against 0.808 in panel B.

This finding deserves emphasis. A \emph{valid} measure of the evidence about calibration is not the same thing as an \emph{optimal} weighting rule. A score that over-penalizes every expert still down-weights the right experts when the miscalibrated ones are clearly worse, and excluding a calibrated generalist costs little when an informed calibrated expert survives. Over-penalization acts like a sharper weighting temperature, which in these designs is mildly beneficial. The correction is therefore best seen as restoring the \emph{interpretation} of the score: as a $p$-value, as the basis for a cut-off, and in comparisons across studies with different seed sets. It is not, by itself, a device to improve point forecasts.

\section{Real data analysis}\label{sec:app}

\subsection{Data}

We use the Classical Model studies distributed with the open-source Anduryl software \citep{thart2019update,rongen2020update}, which reproduce the TU Delft expert judgment database described by \citet{cooke2008tu} and are publicly available \citep{moralesnapoles2019data}. After removing variants of the same study (seed-only subsets, out-of-sample versions and test files) and one study whose realization file is missing, 45 studies remain, with 453 experts and between 7 and 21 seed variables (median 10). Twenty-nine studies elicit three quantiles ($K=4$; 5-50-95 in all but two, which use 10-50-90) and sixteen elicit the 5, 25, 50, 75 and 95\% quantiles ($K=6$). Experts with missing answers on any seed are excluded from that study. Our implementation of~\eqref{eq:T}-\eqref{eq:weights} reproduces the calibration and information scores computed by Anduryl to machine precision on complete-data studies. The analysis has two parts: a database-wide sensitivity analysis, which needs no information on the seeds beyond the elicitation itself, and three case studies in the social sciences whose seeds carry explicit region, period and quantity attributes, so that the dependence can be estimated.

\subsection{How sensitive are published calibration scores to the design effect?}\label{sec:app_db}

\begin{figure}[t]
\centering
\includegraphics[width=\textwidth]{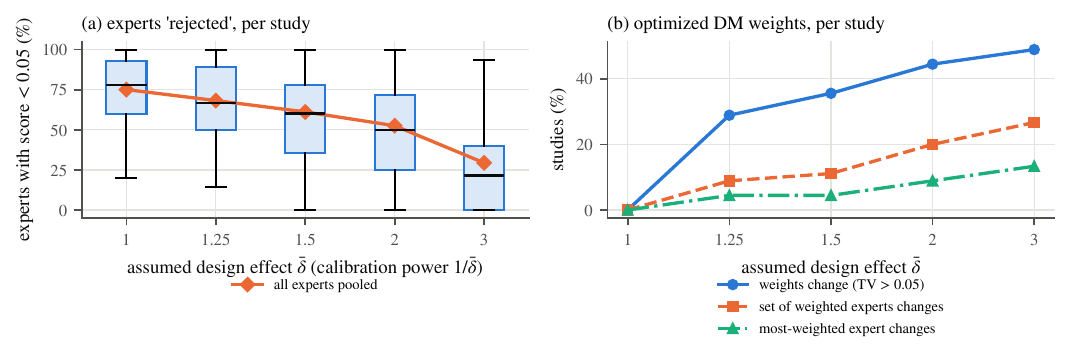}
\caption{Sensitivity of 45 published Classical Model studies (453 experts) to the design effect of the seed set. (a) Per-study share of experts with calibration score below 0.05 (boxes) and share over all experts (line), as a function of the assumed design effect, that is, of the calibration power $1/\bar\delta$. (b) Share of studies in which the optimized global weights change by more than 0.05 in total variation, in which the set of experts with positive weight changes, and in which the most-weighted expert changes.}\label{fig:app_db}
\end{figure}

For each study we recompute every expert's score with an assumed design effect $\bar\delta\in\{1,1.25,1.5,2,3\}$, that is, with calibration power $1/\bar\delta$ in~\eqref{eq:rs1}, and re-optimize the global weights with the DM scored in the same way. Figure~\ref{fig:app_db} summarizes the results. With the classical score, 75.1\% of the 453 experts have a calibration score below 0.05, in line with the well-known finding that most experts are poorly calibrated \citep{cooke2008tu}. The share falls to 61.1\% for $\bar\delta=1.5$ and to 52.5\% for $\bar\delta=2$. The optimized weights are more stable, because the Classical Model typically concentrates the weight on the one or two best-calibrated experts, whose ranking does not change. Even so, a modest design effect of 1.25 moves more than 5\% of the total weight in 28.9\% of the studies, and up to one third of it. With $\bar\delta=2$ the weights move by more than 5\% in 44.4\% of the studies, the set of experts with positive weight changes in 20.0\%, and the most-weighted expert changes in 8.9\%.

\subsection{Three case studies with structured seeds}\label{sec:app_cases}

The three studies concern political violence, terrorism and conflict, and oil markets and economic growth. In all three the seed variables are the same kind of quantity measured in several world regions and periods. Such seeds are exposed to common shocks, for example a global surge of political violence or a commodity cycle, that experts may not have anticipated and that would shift many realizations into the same bins of every expert at once. We place each seed in the question space of Section~\ref{sec:qspace} with three attributes, namely quantity type, region and period, and report the results for the full attribute set and for each attribute alone (Table~\ref{tab:app_struct}), and under alternative range caps (Table~\ref{tab:app_sens}).

\begin{itemize}
\item \emph{Political violence} (21 seeds, 16 experts, of whom 15 answered every seed): counts and shares of terrorist attacks, insurgencies and high-casualty conflicts in eight regions over five periods (the 1970s, 1990s, 2000s, 2010s and 2016). Eight seeds, for example, are the numbers of terrorist attacks in 2016 in each region.
\item \emph{Terrorism and conflict} (18 seeds, 8 experts): a second, independent panel on a closely related set of questions, with thirteen region-by-period seeds of the same kind and five further seeds on conflict-related quantities. It provides an external replication of the first study.
\item \emph{Oil and growth} (20 seeds, 6 experts): average annual GDP growth of a leading economy, of the OECD and of the world in 1991-99, 2002-08 and 2009-12; the OPEC share of world oil demand and the OECD share of imported oil in the same three periods; and proven oil reserves of five producers in 2013.
\end{itemize}

\begin{table}[t]
\centering
\caption{Estimated dependence in the three case studies under alternative question spaces. $\hat\phi$: range (in attribute-mismatch units); $\bar{\hat\nu}$: mean sill share; $\bar{\hat\delta}$: mean (range) of the experts' design effects; $\bar N_{\mathrm{eff}}$: mean effective number of seeds; $\Lambda$: restricted likelihood-ratio statistic against independence, with $p$-value $p_\Lambda$ from 199 parametric-bootstrap replicates under independence.}\label{tab:app_struct}
\footnotesize
\input{tables/tab_app_struct.tex}
\end{table}

\paragraph{Evidence of dependence}
All three studies show significant dependence with the full question space: $\Lambda=60.7$, $20.3$ and $10.1$, with bootstrap $p$-values below 0.01, below 0.01 and 0.025 respectively. The single attributes behave sensibly. Region is a significant source of dependence in all three studies, and quantity type and period matter in the two studies on violence. In the oil study, by contrast, neither quantity type nor period alone shows dependence ($p=0.37$ and $0.75$), whereas region does ($p=0.02$). The procedure therefore does not ``find'' dependence indiscriminately. The estimated design effects with the full question space are 1.64-1.96, 1.00-1.71 and 1.83, so that the seeds carry the information of about 11, 13 and 11 independent seeds instead of 21, 18 and 20. In all three studies the range estimate lies at the identifiability constraint $3\phi\le3$, the diameter of the full question space, and in the political violence and oil studies most estimates $\hat\nu_e$ lie at the upper bound 0.98. The dependence therefore extends over the whole question space, as a common shock would, and its magnitude is partly set by the range constraint rather than by the data.

\paragraph{Sensitivity to the specification}
Table~\ref{tab:app_sens} reports the design effect, and the bootstrap score of the expert who is decisive for the conclusions below, under four specifications: the baseline, a tighter ($4\phi\le3$) and a looser ($2\phi\le3$) range cap, and the single attribute with the largest restricted likelihood. Three findings stand out. First, the upper bound on $\nu_e$ is immaterial: replacing 0.98 by 0.95 or 0.995 changes every mean design effect by less than 0.05, whatever the cap. Second, the range cap is not immaterial. The restricted likelihood increases as the cap is relaxed and $\hat\phi$ stays at the cap, so that the range is not identified beyond it, as anticipated in Section~\ref{sec:estimation}; the mean design effects move from 1.55, 1.24 and 1.47 under the tighter cap to 2.60, 1.86 and 2.57 under the looser one. Third, the data do not single out one question space. In each study the best single attribute fits slightly better than the full set ($\Lambda=63.7$ against 60.7, 25.9 against 20.3, and 11.1 against 10.1). Across the specifications of Table~\ref{tab:app_sens} and the other single attributes with significant dependence, the effective number of seeds ranges from 8.2 to 15.2 out of 21, from 10.3 to 14.8 out of 18, and from 7.8 to 13.6 out of 20. We keep the full attribute set with the cap $3\phi\le3$ as the baseline, because it is the natural description of the seeds, chosen before the analysis, and treats the three attributes symmetrically, and we read the expert-level results below in the light of this sensitivity.

\begin{table}[t]
\centering
\caption{Sensitivity of the case-study results to the range cap and to the question space. Baseline: full attribute set with practical range $3\phi$ at most the diameter, 3, of the question space; $4\phi\le3$ and $2\phi\le3$: tighter and looser caps; last row of each study: the single attribute with the largest $\Lambda$. Columns as in Table~\ref{tab:app_struct}; $C^{\mathrm{PB}}$ is the bootstrap score ($B=9999$) of the expert who is decisive for the conclusions of Section~\ref{sec:app_cases}. The bound $\nu_e\le0.98$ is kept throughout; replacing it by 0.95 or 0.995 changes every mean design effect by less than 0.05.}\label{tab:app_sens}
\footnotesize
\setlength{\tabcolsep}{4pt}
\input{tables/tab_app_sens.tex}
\end{table}

\begin{figure}[t]
\centering
\includegraphics[width=\textwidth]{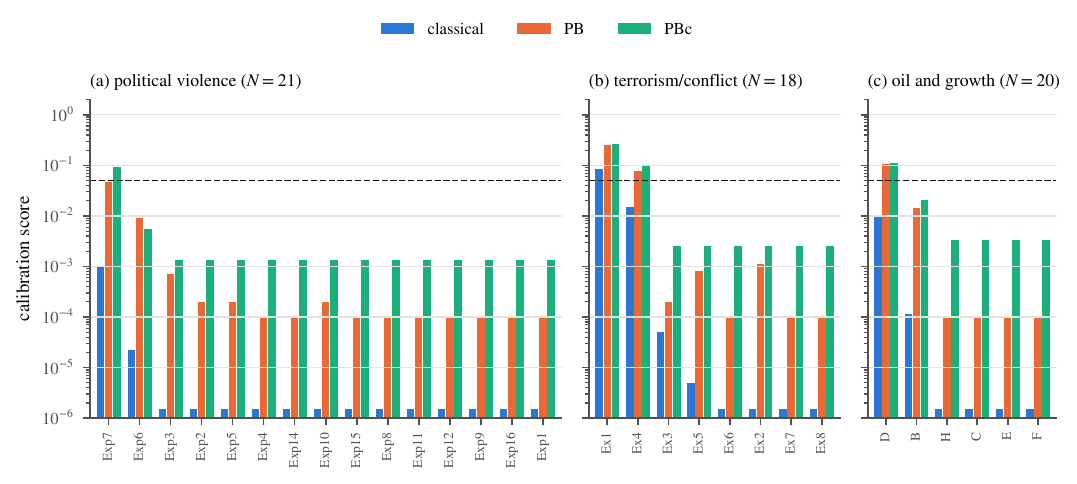}
\caption{Classical, bootstrap-corrected (PB) and prepivoted (PBc) calibration scores in the three case studies, on a logarithmic scale; experts are ordered by their classical score. The dashed line marks 0.05. Scores below the bootstrap resolution are drawn at that resolution.}\label{fig:app_cases}
\end{figure}

\paragraph{Consequences for scores and weights}
Figure~\ref{fig:app_cases} and Tables~\ref{tab:app_pv}-\ref{tab:app_oil} report the scores and weights. The three studies illustrate three different consequences of the correction.

In the political violence study all 15 experts have classical scores below 0.001, and the optimized DM puts 98.2\% of its weight on a single expert (Exp7), with a cut-off of $\alpha^\star=2.3\times10^{-5}$. The correction raises the scores by one to two orders of magnitude but does not rehabilitate the panel. Only Exp7 approaches the conventional threshold ($C^{\mathrm{PB}}=0.047$, $C^{\mathrm{RS2}}=0.052$, and $C^{\mathrm{PBc}}=0.091$ after prepivoting), and the corrected weights shift 13\% of the mass from Exp7 to Exp6 and Exp3. On a common scale the corrected DM also scores higher in-sample: 0.795 against 0.600 when both DMs are scored with the bootstrap calibration, and 0.343 against 0.187 with the classical one. Under the tighter cap or with region alone, however, the bootstrap score of Exp7 is only 0.018 or 0.011 (Table~\ref{tab:app_sens}), so its proximity to the threshold depends on the specification. What holds under every specification is the overall verdict: dependence substantially reduces the effective evidence but does not overturn the conclusion that the panel was poorly calibrated on these seeds.

In the terrorism and conflict study, an expert who is clearly rejected by the classical score (Ex4, $C^{\mathrm{cl}}=0.015$) is no longer rejected once dependence is accounted for ($C^{\mathrm{PB}}=0.079$, $C^{\mathrm{PBc}}=0.097$). The optimized cut-off rises from 0.015 to 0.079, and the weight of Ex4 in the DM rises from 13.8\% to 22.3\%. This reversal is, however, sensitive to the specification: the bootstrap score of Ex4 is 0.043 under the tighter cap and 0.017 with quantity type alone (Table~\ref{tab:app_sens}), so it is a sensitivity result rather than a firm rehabilitation. Nor do the corrected weights improve the DM in-sample. Scored with the bootstrap calibration, the DM with classical weights has a combined score of 0.931 and the DM with corrected weights 0.643 (0.660 and 0.386 with the classical calibration). The replication is consistent with the first study: in both panels the estimated dependence is substantial, and the correction changes the weights without changing which experts dominate.

In the oil and growth study, the classical analysis gives all the weight to a single expert (D, $C^{\mathrm{cl}}=0.011$), with all others below 0.001, and its DM is itself poorly calibrated: its in-sample combined score is 0.010. With the corrected scores, expert D rises to $C^{\mathrm{PB}}=0.107$, above the conventional threshold, and a second expert (B, $C^{\mathrm{PB}}=0.014$) enters the DM with 16.5\% of the weight. Expert D stays above 0.05 under every specification of Table~\ref{tab:app_sens} (0.057-0.212), whereas the entry of B depends on the specification: with region alone its score is 0.003 and D again receives all the weight. The two DMs must be compared on a common scale, because each is optimized and scored with its own calibration score. Scored with the bootstrap calibration, the corrected DM has a combined score of 0.232 against 0.096 for the classical one; scored with the classical calibration, 0.101 against 0.010. Here the correction changes the composition of the decision maker and, in-sample, improves it on either scale.

\begin{table}[t]
\centering
\caption{Political violence study ($N=21$ seeds, 15 experts): calibration statistic, estimated design effect and effective number of seeds, information score, classical and corrected calibration scores, and optimized global weights under the classical and bootstrap scores. $C^{\mathrm{PB}}$ uses $B=9999$; $C^{\mathrm{PBc}}$ is prepivoted with $R=49$ and $B=199$, so its resolution is $1/736$. Optimized cut-off: $\alpha^\star=2.3\times10^{-5}$ (classical) and $7\times10^{-4}$ (PB); in-sample combined score of the DM with classical and with PB weights: 0.187 and 0.343 when the DM is scored with the classical calibration, 0.600 and 0.795 when it is scored with the bootstrap calibration ($B=99{,}999$).}\label{tab:app_pv}
\scriptsize
\setlength{\tabcolsep}{3.6pt}
\input{tables/tab_app_pv.tex}
\end{table}

\begin{table}[t]
\centering
\caption{Terrorism and conflict study ($N=18$ seeds, 8 experts); columns as in Table~\ref{tab:app_pv}. The resolution of $C^{\mathrm{PBc}}$ is $1/393$. Optimized cut-off: $\alpha^\star=0.015$ (classical) and $0.079$ (PB); in-sample combined score of the DM with classical and with PB weights: 0.660 and 0.386 (classical calibration), 0.931 and 0.643 (bootstrap calibration).}\label{tab:app_rav}
\scriptsize
\setlength{\tabcolsep}{3.6pt}
\input{tables/tab_app_rav.tex}
\end{table}

\begin{table}[t]
\centering
\caption{Oil and growth study ($N=20$ seeds, 6 experts); columns as in Table~\ref{tab:app_pv}. The resolution of $C^{\mathrm{PBc}}$ is $1/295$. Optimized cut-off: $\alpha^\star=0.011$ (classical) and $0.014$ (PB); in-sample combined score of the DM with classical and with PB weights: 0.010 and 0.101 (classical calibration), 0.096 and 0.232 (bootstrap calibration).}\label{tab:app_oil}
\scriptsize
\setlength{\tabcolsep}{3.6pt}
\input{tables/tab_app_oil.tex}
\end{table}

Across the three studies the corrected analysis tells a consistent story at the level of the seed set and a more qualified one at the level of individual experts. Seeds that measure the same kind of quantity across regions and periods carry between about one half and three quarters of their nominal information under the baseline specification, and between about two fifths and four fifths across the specifications considered, so the classical score overstates the evidence against every expert. Correcting for this does not reverse the overall verdict on any panel. It changes cut-offs and weights, and it can lift individual experts above conventional thresholds and change the composition of the DM; whether it does so for a given expert depends on how the dependence is specified, which is why the sensitivity analysis should be reported alongside the corrected scores.
\section{Discussion and Conclusions}\label{sec:discussion}

The calibration score is usually read as a property of the expert. Our analysis shows that, as a number, it is also a property of the seed set, namely of how many effectively independent pieces of evidence the seeds contain. Two panels assessed on equally relevant seeds, one drawn from a single indicator in several regions and one drawn from unrelated domains, produce calibration scores that are not on the same scale. The procedures guide already warns that scores based on different numbers of realizations cannot be compared directly \citep{cooke2000procedures}; the design effect $\dbar$ extends this warning to seed sets of equal size but different dependence. The database analysis of Section~\ref{sec:app_db} shows that the share of experts deemed poorly calibrated, and in a sizeable minority of studies the weights themselves, are sensitive to design effects as small as 1.25-1.5. Reporting $\hat N_{\mathrm{eff}}$ next to $N$ in every Classical Model study would be a cheap and informative first step, even when the scores themselves are left uncorrected.

Because the first-order correction coincides with the calibration power of existing software, practitioners can apply it without new software by setting the power to $1/\hat{\bar\delta}$. The difference from current practice is that the power becomes an \emph{estimated nuisance parameter}, reported with its basis, rather than a free constant that might be tuned to obtain a desired result. The second-order and bootstrap scores are preferable when the eigenvalues are unequal, as they generically are for the Cooke format under dependence (Figure~\ref{fig:deff}c), and at small $N$, where the discreteness of $T$ already distorts the classical score under independence. For non-geographical seeds, the dependence structure must be specified through attributes, and different analysts may choose different attributes. We recommend reporting the results for the natural attribute set together with each attribute alone, as in Table~\ref{tab:app_struct}, and using the restricted likelihood-ratio check to decide whether a correction is needed. The range cap deserves the same treatment: in our case studies it binds, and reporting the results under a tighter and a looser cap, as in Table~\ref{tab:app_sens}, shows how much of the correction rests on it. In the case studies this check distinguishes between attributes that carry dependence and attributes that do not, such as period and quantity type in the oil study; and under independence, in the simulations, the correction is nearly harmless (Table~\ref{tab:size}). Seeds should ideally be designed with this in mind: recording the region, period and quantity of each seed costs nothing at the design stage and makes the dependence estimable later.

The main limitation is intrinsic rather than methodological. All experts are scored against one set of realizations, and the component of that set that most affects the calibration statistic, the common deviation of the seeds, is the one about which the data are least informative, because it is confounded with a common bias of the expert. The REML estimator on rank normal scores is invariant to such biases and approximately median-unbiased for the design effect, and prepivoting propagates its variability into the scores (Section~\ref{sec:sim_prepivot}). Two forms of dependence nevertheless remain beyond the reach of any procedure that uses the elicitation data alone. The first is dependence whose range exceeds the seed set, which is indistinguishable from a constant shift. The second is dependence acting through a common scale, which is indistinguishable from common over- or under-confidence (Section~\ref{sec:robust}). Both can be addressed only with external information, such as historical replicates of the seed variables, which are often available for environmental and economic seeds and would identify the correlation function independently of the elicitation.

There is a long-running debate on whether performance weights outperform equal weights out of sample \citep{clemen2008comment,flandoli2011comparison,colson2017cross,cooke2021expert,marti2021performance}. Our results add two considerations to it. When seeds are dependent, the evidence that separates experts is weaker than the classical scores suggest, and uninformed experts are penalized more than informed ones for reasons unrelated to their calibration. Yet, in our simulations, the stronger discrimination of the classical score did not harm the DM, because the experts it penalized were mostly genuinely miscalibrated. Whether dependence-robust scores improve DMs in real panels is therefore an empirical question, and the effective number of seeds is a natural covariate in meta-analyses comparing performance and equal weighting. Our results also bear on cross-validation \citep{colson2017cross}. Random splits of dependent seeds into training and test sets share the same common deviation, so they reward experts who were right about \emph{these} realizations rather than calibrated experts; splits that keep related seeds together would be the appropriate alternative.

The same theory applies to the Delphi-style and IDEA-style protocols that use seed questions to score participants \citep{hemming2018practical,rowe1999delphi}, whose small-sample ordinal data raise related problems of dependence among items \citep{calleo2026community}, and to calibration checks of probabilistic forecasts expressed as quantiles \citep{gneiting2007probabilistic}. For clustered seeds with within-cluster latent correlation $\rho$ and cluster sizes $n_c$, $\kappa_{N,m}=\rho^m N^{-1}\sum_cn_c(n_c-1)$, and the design effect follows immediately from~\eqref{eq:deff}; this is the multinomial counterpart of the threshold-specific design effect of \citet{noonan2026exceedance}. In time series, heteroskedasticity- and autocorrelation-consistent estimators of $\SigN$ are non-parametric alternatives to the copula model, as are spatial kernel estimators \citep{conley1999gmm}; with the 10-30 seeds typical of SEJ studies, however, they are very noisy. A fully Bayesian treatment, with a prior on $(\phi,\bm\nu)$ and the posterior predictive $p$-value of $T$ as the score, would integrate over the dependence rather than plug it in, and is natural when prior knowledge about the seed variables is available.

The calibration score of Cooke's Classical Model compares an expert's bin frequencies with a $\chi^2$ distribution that is correct only when the seed variables are independent. When seeds share a common driver, the statistic follows a weighted $\chi^2$ law, the classical score is anti-conservative, and calibrated experts, especially uninformed ones, are penalized. A Mehler-Hermite expansion makes the long-run covariance of the bin indicators explicit, defines the design effect of the seed set and the effective number of seed variables, and shows that the first-order correction is precisely the calibration power already available in practice. Rao-Scott and parametric-bootstrap corrections, combined with a REML estimator of the dependence that is invariant to miscalibration and with prepivoting, largely restore the nominal behavior of the score, with limits that we have characterized. In published studies, scores and, in a sizeable minority of cases, weights are sensitive to modest design effects, and in three studies with structured seeds the effective number of seeds is between about one half and three quarters of the nominal one under the baseline specification, with conclusions about individual experts that depend on how the dependence is specified. We recommend that applications of the Classical Model record the attributes of their seeds, report the effective number of seed variables, and, where the seeds share a common source, report dependence-robust scores and the weights derived from them as a sensitivity analysis.

\section*{Software and data availability}
The expert judgment data are publicly available \citep{moralesnapoles2019data} and are distributed with the Anduryl software. Python code implementing the methods (bivariate normal probabilities, Hermite design effects, corrected scores, REML and pairwise-likelihood estimators, prepivoted bootstrap and the Classical Model weighting) and reproducing every table and figure is provided as Supplementary material S1, with a README describing each script. The file is \texttt{supplementary\_code.zip}.

\section*{Funding}
This research did not receive any specific grant from funding agencies in the public, commercial, or not-for-profit sectors.

\section*{Declaration of competing interest}
The author declares no known competing financial interests or personal relationships that could have appeared to influence the work reported in this paper.

\section*{Declaration of generative AI and AI-assisted technologies in the manuscript preparation process}
During the preparation of this work, the author used Claude (Anthropic) to review the language and ensure internal consistency of the text. After using this tool, the author reviewed and edited the content as needed and takes full responsibility for the content of the published article.

\bibliographystyle{elsarticle-harv}
\bibliography{references}
\appendix
\setcounter{table}{0}
\setcounter{figure}{0}

\section{Derivations}\label{app:proofs}

\subsection{The limit law~\eqref{eq:limit}}\label{app:limit}

\emph{Central limit theorem.} Fix $\bm a\in\mathbb R^K$ and let $Y_i=\bm a^\top(\bZ_i-\bp)$, so that $|Y_i|\le 2\|\bm a\|_\infty$ and $\E Y_i=0$ under $H_0$. The variance $\Var(N^{-1/2}\sum_iY_i)=\bm a^\top\SigN\bm a$ converges to $\bm a^\top\bm\Sigma\bm a$. If $\bm a^\top\bm\Sigma\bm a>0$, the central limit theorem for bounded $\alpha$-mixing arrays of random fields \citep[Theorem~1]{jenish2009central} gives $N^{-1/2}\sum_iY_i\xrightarrow{d}\mathcal N(0,\bm a^\top\bm\Sigma\bm a)$; if $\bm a^\top\bm\Sigma\bm a=0$, the variance tends to zero and $N^{-1/2}\sum_iY_i\xrightarrow{p}0$. By the Cram\'er-Wold device, $\sqrt N(\bs-\bp)\xrightarrow{d}\mathcal N_K(\bm 0,\bm\Sigma)$, and in particular $\bs\xrightarrow{p}\bp$.

\emph{Expansion.} Let $f(\bs)=\sum_ks_k\log(s_k/p_k)$. On the event $\mathcal E_N=\{\min_ks_k>\min_kp_k/2\}$, whose probability tends to one, $f$ is three times continuously differentiable with bounded third derivatives, with $f(\bp)=0$, $\partial f/\partial s_k|_{\bp}=1$ and $\partial^2f/\partial s_k\partial s_l|_{\bp}=\delta_{kl}/p_k$. A Taylor expansion with Lagrange remainder gives $f(\bs)=\one^\top(\bs-\bp)+\tfrac12(\bs-\bp)^\top\Dp^{-1}(\bs-\bp)+R_N$ with $|R_N|\le C\|\bs-\bp\|^3$ on $\mathcal E_N$. Since $\one^\top\bs=\one^\top\bp=1$, the linear term vanishes, and $T=2Nf(\bs)=\bm Y_N^\top\bm Y_N+O_p(N^{-1/2})$ with $\bm Y_N=\sqrt N\Dp^{-1/2}(\bs-\bp)$.

\emph{Quadratic form.} $\bm Y_N\xrightarrow{d}\bm G\sim\mathcal N_K(\bm 0,\bm A)$, hence $T\xrightarrow{d}\bm G^\top\bm G$ by Slutsky's lemma and the continuous mapping theorem. Writing $\bm A=\sum_k\lambda_k\bm v_k\bm v_k^\top$ with orthonormal eigenvectors, $\bm G^\top\bm G=\sum_{k:\lambda_k>0}\lambda_k\xi_k^2$ with $\xi_k=\bm v_k^\top\bm G/\sqrt{\lambda_k}$ i.i.d.\ standard normal. Because $\bm A\sqrt{\bp}=\Dp^{-1/2}\bm\Sigma\one=\bm 0$ and $\bm\Sigma$ has rank $K-1$, exactly $K-1$ eigenvalues are positive. Finally, $Q_{\bm\lambda}\overset{d}{=}\chi^2_{K-1}$ forces $\sum_k\lambda_k=K-1$ and $\sum_k\lambda_k^2=K-1$ (equal means and variances), hence $\sum_k(\lambda_k-1)^2=0$, which holds if and only if $\bm A=\bm I-\sqrt{\bp}\sqrt{\bp}^\top$, that is, $\bm\Sigma=\Dp-\bp\bp^\top$.

\subsection{The direction of the distortion~\eqref{eq:anticons}}\label{app:anticons}

Let $\bm\Delta=\SigN-(\Dp-\bp\bp^\top)\succeq0$. Since both $\SigN$ and $\Dp-\bp\bp^\top$ annihilate $\one$, so does $\bm\Delta$. Then $\bm A_N=(\bm I-\sqrt{\bp}\sqrt{\bp}^\top)+\tilde{\bm\Delta}$ with $\tilde{\bm\Delta}=\Dp^{-1/2}\bm\Delta\Dp^{-1/2}\succeq0$ and $\tilde{\bm\Delta}\sqrt{\bp}=\bm 0$. On the invariant subspace orthogonal to $\sqrt{\bp}$, $\bm A_N=\bm I+\tilde{\bm\Delta}$, whose eigenvalues are at least one by the Courant-Fischer theorem. In the limit, $Q_{\bm\lambda}-\sum_k\xi_k^2=\sum_k(\lambda_k-1)\xi_k^2\ge0$ almost surely, so with $c=\chi^2_{K-1,1-\alpha}$ we have $\Pr(Q_{\bm\lambda}\ge c)\ge\Pr(\chi^2_{K-1}\ge c)=\alpha$, with strict inequality if some $\lambda_k>1$ because the event $\{\sum_k\xi_k^2<c\le Q_{\bm\lambda}\}$ then has positive probability. Since $c$ is a continuity point of the law of $Q_{\bm\lambda}$, \eqref{eq:limit} gives $\Pr(C^{\mathrm{cl}}\le\alpha)=\Pr(T\ge c)\to\Pr(Q_{\bm\lambda}\ge c)$.

\subsection{The Hermite representation~\eqref{eq:covpair}-\eqref{eq:SigmaHermite}}\label{app:hermite}

We use Mehler's formula in the form of \citet{lancaster1958structure}: if $(W,W')$ is standard bivariate normal with correlation $|\rho|<1$ and $f,g\in L^2(\varphi)$ have Hermite coefficients $\hat f_m=\E[f(W)\He_m(W)]$, then $\E[f(W)g(W')]=\sum_{m\ge0}\rho^m\hat f_m\hat g_m/m!$. The series converges absolutely, because by the Cauchy-Schwarz inequality and Parseval's identity $\sum_m\hat f_m^2/m!=\E f(W)^2$ (recall that $\E\He_m(W)^2=m!$) we have $\sum_m|\rho|^m|\hat f_m\hat g_m|/m!\le\|f\|\,\|g\|$. Take $f=\mathbb 1_{B_a}$ and $g=\mathbb 1_{B_b}$ with $B_k=(\tau_{k-1},\tau_k]$. Then $\hat f_0=p_a$, and for $m\ge1$ the identity $\frac{d}{dx}\{\He_{m-1}(x)\varphi(x)\}=-\He_m(x)\varphi(x)$, which follows from $\He_{m-1}'=(m-1)\He_{m-2}$ and the three-term recurrence, gives
\[
\hat f_m=\int_{\tau_{a-1}}^{\tau_a}\He_m(x)\varphi(x)\,dx=\varphi(\tau_{a-1})\He_{m-1}(\tau_{a-1})-\varphi(\tau_a)\He_{m-1}(\tau_a)=c_{m,a},
\]
the boundary terms at $\pm\infty$ vanishing. Subtracting the $m=0$ term $p_ap_b$ yields~\eqref{eq:covpair}; summing over the finitely many pairs $i\neq j$ and adding the diagonal contribution $\Var(\bZ_i)=\Dp-\bp\bp^\top$ yields~\eqref{eq:SigmaHermite}. The sum $\sum_kc_{m,k}$ telescopes to $\varphi(\tau_0)\He_{m-1}(\tau_0)-\varphi(\tau_K)\He_{m-1}(\tau_K)=0$, and Parseval's identity for $\mathbb 1_{B_a}$ and $\mathbb 1_{B_b}$ (by polarization) gives $\sum_{m\ge1}c_{m,a}c_{m,b}/m!=\delta_{ab}p_a-p_ap_b$.

\subsection{The design effect~\eqref{eq:deff} and its bounds~\eqref{eq:bounds}}\label{app:deff}

Multiplying~\eqref{eq:SigmaHermite} by $\Dp^{-1}$ and taking traces, with $\tr\{\Dp^{-1}(\Dp-\bp\bp^\top)\}=K-\sum_kp_k=K-1$, gives $\tr(\Dp^{-1}\SigN)=K-1+\sum_mg_m\kappa_{N,m}$, which is~\eqref{eq:deff}. The same operation applied to $\sum_{m\ge1}\bc_m\bc_m^\top/m!=\Dp-\bp\bp^\top$ gives $\sum_mg_m=K-1$. If $0\le\rho_{ij}<1$, every $\kappa_{N,m}$ is non-negative, so truncating the series gives the lower bound in~\eqref{eq:bounds}. Moreover $\rho_{ij}^m\le\rho_{ij}$, so $\kappa_{N,m}\le\kappa_{N,1}$ and $\sum_mg_m\kappa_{N,m}\le(K-1)\kappa_{N,1}$, which is the upper bound. The matrix $\sum_m(\kappa_{N,m}/m!)\bc_m\bc_m^\top$ is positive semi-definite, so the argument of \ref{app:anticons} gives $\lambda_{N,k}\ge1$. Finally, $\Var(\sqrt N\bar W)=N^{-1}\sum_{i,j}\rho_{ij}=1+\kappa_{N,1}$.

\subsection{The ceiling~\eqref{eq:ceiling}}\label{app:ceiling}

If $\rho_{ij}\ge\rho_{\min}$ for all $i\neq j$, then $\kappa_{N,1}\ge(N-1)\rho_{\min}$, and the lower bound~\eqref{eq:bounds} with $M=1$ gives $\dbar\ge1+\omega(N-1)$ with $\omega=g_1\rho_{\min}/(K-1)$. Some $g_m$ with $m\ge2$ is positive: otherwise $\bc_m=\bm 0$ for every $m\ge2$, and each bin indicator would equal, in $L^2(\varphi)$, an affine function of $W$, which is impossible for an indicator. For Cooke's format, $g_2=0.6395$. Since $\sum_mg_m=K-1$, we have $g_1<K-1$ and hence $0<\omega<\rho_{\min}\le1$. Then $\Neff\le N/\{1+\omega(N-1)\}$, and $N/\{1+\omega(N-1)\}<1/\omega$ is equivalent to $\omega<1$.

\subsection{Validity of plug-in scores}\label{app:valid}

The distribution function $F_{\bm\lambda}(x)=\Pr(Q_{\bm\lambda}\le x)$ is jointly continuous in $(x,\bm\lambda)$ on $[0,\infty)\times(0,\infty)^{K-1}$. If $\hat{\bm\lambda}\xrightarrow{p}\bm\lambda$, as under increasing-domain asymptotics in which $(\hat\phi,\hat{\bm\nu})$ is consistent, then $(T,\hat{\bm\lambda})\xrightarrow{d}(Q_{\bm\lambda},\bm\lambda)$ by~\eqref{eq:limit} and Slutsky's lemma, and the continuous mapping theorem gives $1-F_{\hat{\bm\lambda}}(T)\xrightarrow{d}1-F_{\bm\lambda}(Q_{\bm\lambda})\sim\mathrm{Uniform}(0,1)$. The same argument applies to $C^{\mathrm{RS1}}$ and $C^{\mathrm{RS2}}$, which are continuous in $(T,\hat{\bm\lambda})$. The limit of $C^{\mathrm{RS1}}$ is uniform if and only if $Q_{\bm\lambda}/\dbar\sim\chi^2_{K-1}$; the means always agree, and the variances agree if and only if $\sum_k\lambda_k^2=(K-1)\dbar^2$, that is, by the Cauchy-Schwarz inequality, if and only if all $\lambda_k$ are equal.

\subsection{Invariance of the dependence estimator}\label{app:inv}

Suppose that for expert $e$ the PITs satisfy $\tilde U_{ei}=G_e(W_{ei})$ for a strictly increasing function $G_e$ common to her seeds. Replacing $W_{ei}$ by $(W_{ei}-b_e)/\sigma_e$, with $b_e\in\mathbb R$ and $\sigma_e>0$, replaces $\tilde U_{ei}$ by $H_e(\tilde U_{ei})$ with $H_e(x)=G_e\{(G_e^{-1}(x)-b_e)/\sigma_e\}$, which is strictly increasing. Ranks are invariant under strictly increasing maps, so the rank normal scores and the estimate maximizing~\eqref{eq:reml} are unchanged. For piecewise-uniform PITs the condition holds exactly within the central bins; in the tail bins it holds up to the item-specific intrinsic range, which affects only the ordering of realizations that fall in the same tail bin.
\newpage
\section{Computational details}\label{app:comp}

\paragraph{Bivariate normal probabilities}
All rectangle probabilities of~\eqref{eq:copula} are obtained from
\[
\Phi_2(h,k;\rho)=\Phi(h)\Phi(k)+\frac1{2\pi}\int_0^{\arcsin\rho}\exp\Big\{-\frac{h^2-2hk\sin\theta+k^2}{2\cos^2\theta}\Big\}d\theta ,
\]
which follows from Plackett's identity $\partial\Phi_2/\partial\rho=\varphi_2(h,k;\rho)$ and the substitution $\rho=\sin\theta$, which removes the singularity at $|\rho|=1$. The integral is evaluated by 32-point Gauss-Legendre quadrature; against an independent implementation, the maximum absolute error over 400 random cases with $|\rho|\le0.99$ was $1.7\times10^{-16}$. For the long-run covariance, the cell covariances are tabulated on a grid of 1601 values of $\rho\in[-0.995,0.995]$ and interpolated linearly, with maximum error below $1.5\times10^{-6}$ in $\SigN$.

\paragraph{Numerical checks}
The closed form of $\bc_m$ agreed with direct quadrature to $2.2\times10^{-16}$ for $m\le5$, and the 60-term Hermite series reproduced the exact cell covariances to $6\times10^{-14}$ at $\rho=0.7$. With $N=25$ random seeds and $(\nu,\phi)=(0.8,0.2)$, the exact $\SigN$ agreed with the 80-term series to $9\times10^{-15}$ and with a Monte Carlo estimate from $2\times10^5$ replicates to $7\times10^{-4}$. Imhof probabilities agreed with Monte Carlo to the third decimal. In an increasing-domain experiment with $N=400$ seeds, the empirical 0.5, 0.9, 0.95 and 0.99 quantiles of $T$ were $(5.26,15.75,20.87,33.39)$, against $(5.22,15.96,21.05,33.75)$ for the weighted-$\chi^2$ limit~\eqref{eq:limit} and $(2.37,6.25,7.81,11.34)$ for $\chi^2_3$.

\paragraph{Optimization}
The restricted likelihood~\eqref{eq:reml} is profiled over $\phi$ on 30 log-spaced values between a small lower bound and $\phi_{\max}$, one third of the diameter of the seed set (Section~\ref{sec:estimation}); for each $\phi$, every $\nu_e$ is maximized over 50 values in $[0,0.98]$ (Section~\ref{sec:app_cases} shows that this bound is immaterial in the case studies), and the search is then refined on a $9\times99$ grid around the optimum. One Cholesky factorization per $(\phi,\nu)$ serves all experts. With $E=6$ and $N=100$ a fit takes about 0.4\,s in vectorized Python.

\paragraph{Data processing}
Realizations and quantiles coded as $-999.5$ or $-999.6$ in the Excalibur files are treated as missing. Items on a logarithmic background are scored on the log scale, as in Excalibur. The calibration and information scores computed by our implementation coincide with those of Anduryl to machine precision for studies without missing answers; in the studies with incomplete answers that we checked, small differences arise because we exclude experts with missing seeds.
\newpage
\section{Additional simulation results}\label{app:extra}

\begin{table}[h]
\centering
\caption{As Table~\ref{tab:size}, at the 1\% level. Classical rates above 1.5\% are in bold.}\label{tab:size01}
\footnotesize
\input{tables/tab_size_01.tex}
\end{table}
\end{document}

%% file: tables/tab_gm.tex
\begin{tabular}{lcccccccc}
\toprule
$m$ & 1 & 2 & 3 & 4 & 5 & 6 & 7 & 8 \\
\midrule
$g_m$ & 0.8144 & 0.6395 & 0.4511 & 0.0046 & 0.2449 & 0.0398 & 0.0761 & 0.0647 \\
$\sum_{l\le m} g_l/(K-1)$ & 0.271 & 0.485 & 0.635 & 0.637 & 0.718 & 0.731 & 0.757 & 0.778 \\
\bottomrule
\end{tabular}

%% file: tables/tab_size.tex
\begin{tabular}{rr c ccccccc}
\toprule
$N$ & $\phi$ & $\bar\delta_N$ & $\widehat{\bar\delta}_N$ (med.) & Classical & RS1$^{\ast}$ & PB$^{\ast}$ & RS1 & RS2 & PB \\
\midrule
10 & 0.00 & 1.00 & 1.00 & 2.9 & 2.9 & 4.1 & 2.0 & 2.0 & 2.8 \\
 & 0.05 & 1.03 & 1.09 & 3.6 & 3.4 & 4.3 & 2.4 & 2.3 & 3.0 \\
 & 0.10 & 1.12 & 1.16 & 5.6 & 4.2 & 4.8 & 3.4 & 3.3 & 4.0 \\
 & 0.20 & 1.36 & 1.36 & \textbf{10.3} & 4.9 & 5.2 & 5.2 & 5.0 & 5.9 \\
 & 0.30 & 1.62 & 1.47 & \textbf{14.3} & 5.0 & 4.4 & 7.8 & 7.5 & 8.9 \\
\addlinespace
20 & 0.00 & 1.00 & 1.00 & 4.3 & 4.3 & 5.3 & 3.2 & 3.2 & 3.9 \\
 & 0.05 & 1.07 & 1.11 & 4.4 & 4.1 & 4.0 & 2.8 & 2.7 & 3.0 \\
 & 0.10 & 1.24 & 1.30 & \textbf{8.1} & 3.8 & 4.5 & 4.0 & 3.9 & 4.2 \\
 & 0.20 & 1.75 & 1.69 & \textbf{17.8} & 4.8 & 4.4 & 7.1 & 6.7 & 7.1 \\
 & 0.30 & 2.32 & 1.99 & \textbf{24.8} & 4.5 & 4.0 & 10.2 & 9.6 & 10.3 \\
\addlinespace
30 & 0.00 & 1.00 & 1.00 & 5.2 & 5.2 & 5.2 & 4.2 & 4.2 & 3.9 \\
 & 0.05 & 1.11 & 1.14 & 6.6 & 4.5 & 4.6 & 4.1 & 4.0 & 3.9 \\
 & 0.10 & 1.38 & 1.39 & \textbf{12.8} & 4.7 & 4.6 & 5.5 & 5.3 & 5.2 \\
 & 0.20 & 2.16 & 2.02 & \textbf{25.8} & 5.2 & 4.3 & 8.0 & 7.3 & 7.7 \\
 & 0.30 & 3.02 & 2.59 & \textbf{39.1} & 4.8 & 3.9 & 11.8 & 10.6 & 10.8 \\
\addlinespace
50 & 0.00 & 1.00 & 1.00 & 5.9 & 5.9 & 4.8 & 5.2 & 5.2 & 4.1 \\
 & 0.05 & 1.18 & 1.21 & \textbf{9.4} & 5.4 & 4.6 & 5.3 & 5.2 & 4.2 \\
 & 0.10 & 1.64 & 1.62 & \textbf{18.7} & 5.4 & 4.8 & 6.6 & 6.1 & 6.0 \\
 & 0.20 & 2.95 & 2.79 & \textbf{40.3} & 4.8 & 3.6 & 10.0 & 8.7 & 8.7 \\
 & 0.30 & 4.40 & 3.79 & \textbf{57.1} & 5.6 & 4.4 & 14.7 & 12.7 & 12.9 \\
\addlinespace
100 & 0.00 & 1.00 & 1.00 & 6.1 & 6.1 & 5.3 & 5.2 & 5.2 & 4.7 \\
 & 0.05 & 1.36 & 1.35 & \textbf{13.6} & 7.1 & 6.0 & 7.3 & 6.8 & 5.7 \\
 & 0.10 & 2.28 & 2.27 & \textbf{29.0} & 6.3 & 4.6 & 7.2 & 5.8 & 5.6 \\
 & 0.20 & 4.96 & 4.47 & \textbf{62.3} & 7.7 & 5.0 & 13.4 & 11.4 & 11.4 \\
 & 0.30 & 7.86 & 7.35 & \textbf{68.4} & 4.8 & 3.4 & 11.7 & 8.9 & 8.9 \\
\bottomrule
\end{tabular}

%% file: tables/tab_prepivot.tex
\begin{tabular}{rr ccc}
\toprule
$N$ & $\phi$ & Classical & PB & PBc \\
\midrule
30 & 0.0 & 4.5 & 2.8 & 3.2 \\
30 & 0.1 & 11.8 & 5.2 & 6.1 \\
30 & 0.2 & 26.1 & 10.1 & 8.3 \\
30 & 0.3 & 36.2 & 10.1 & 8.6 \\
50 & 0.3 & 52.9 & 14.1 & 10.5 \\
\bottomrule
\end{tabular}

%% file: tables/tab_robust.tex
\begin{tabular}{l ccc ccc ccc}
\toprule
 & \multicolumn{3}{c}{$\phi=0.1$} & \multicolumn{3}{c}{$\phi=0.2$} & \multicolumn{3}{c}{$\phi=0.3$} \\
\cmidrule(lr){2-4}\cmidrule(lr){5-7}\cmidrule(lr){8-10}
True copula & Cl. & RS2 & PB & Cl. & RS2 & PB & Cl. & RS2 & PB \\
\midrule
Gaussian (Table~\ref{tab:size}) & 12.8 & 5.3 & 5.2 & 25.8 & 7.3 & 7.7 & 39.1 & 10.6 & 10.8 \\
Non-central $\chi^2_1$ & 9.3 & 4.3 & 4.3 & 22.0 & 8.9 & 8.8 & 31.8 & 12.2 & 12.9 \\
Student $t_4$, common scale & 28.5 & 14.4 & 14.5 & 43.0 & 15.5 & 15.7 & 51.7 & 19.7 & 20.1 \\
\bottomrule
\end{tabular}

%% file: tables/tab_power.tex
\begin{tabular}{ll ccc cccc cccc}
\toprule
 & & \multicolumn{3}{c}{$\phi=0$} & \multicolumn{4}{c}{$\phi=0.1$} & \multicolumn{4}{c}{$\phi=0.2$}\\
\cmidrule(lr){3-5}\cmidrule(lr){6-9}\cmidrule(lr){10-13}
$b$ & $\sigma$ & Cl. & RS1 & PB & Cl. & RS1$^\ast$ & RS1 & PB & Cl. & RS1$^\ast$ & RS1 & PB \\
\midrule
0.00 & 1.00 & 4.3 & 3.2 & 3.4 & 14.4 & 6.0 & 6.4 & 6.4 & 27.8 & 4.9 & 9.0 & 8.6 \\
0.25 & 1.00 & 16.5 & 13.2 & 12.5 & 24.7 & 13.7 & 11.9 & 11.7 & 36.6 & 10.7 & 14.7 & 14.0 \\
0.50 & 1.00 & 53.8 & 47.4 & 45.5 & 55.3 & 39.5 & 35.5 & 34.7 & 58.2 & 26.0 & 28.3 & 27.5 \\
0.75 & 1.00 & 90.0 & 84.3 & 83.5 & 82.5 & 69.6 & 61.2 & 59.6 & 77.9 & 48.1 & 48.1 & 46.2 \\
1.00 & 1.00 & 99.2 & 97.9 & 97.2 & 96.5 & 91.8 & 86.4 & 85.5 & 92.1 & 71.1 & 70.1 & 68.6 \\
\addlinespace
0.00 & 0.80 & 22.9 & 19.1 & 18.0 & 32.2 & 17.6 & 15.9 & 15.5 & 42.7 & 13.0 & 14.0 & 13.4 \\
0.00 & 0.65 & 65.0 & 58.6 & 57.9 & 68.5 & 51.0 & 43.8 & 43.3 & 71.5 & 33.7 & 35.4 & 33.6 \\
0.00 & 0.50 & 95.6 & 92.9 & 92.4 & 95.2 & 89.4 & 85.0 & 83.9 & 94.5 & 73.2 & 73.2 & 71.0 \\
\bottomrule
\end{tabular}

%% file: tables/tab_weights.tex
\begin{tabular}{ll ccc ccc}
\toprule
 & & \multicolumn{3}{c}{Panel A (half miscalibrated)} & \multicolumn{3}{c}{Panel B (mostly calibrated)} \\
\cmidrule(lr){3-5}\cmidrule(lr){6-8}
$\phi$ & Scheme & $w_{\mathrm{mis}}$ & $E_{\mathrm{eff}}$ & Rel.\ QS & $w_{\mathrm{mis}}$ & $E_{\mathrm{eff}}$ & Rel.\ QS \\
\midrule
0 & Classical, optimised & 0.097 & 1.40 & 0.952 & 0.164 & 1.29 & 1.013 \\
 & PB, optimised & 0.105 & 1.33 & 0.961 & 0.164 & 1.22 & 1.012 \\
 & Classical, $\alpha=0$ & 0.133 & 3.11 & 0.830 & 0.119 & 4.44 & 0.840 \\
 & PB, $\alpha=0$ & 0.147 & 3.24 & 0.831 & 0.121 & 4.62 & 0.839 \\
\addlinespace
0.1 & Classical, optimised & 0.101 & 1.36 & 0.940 & 0.145 & 1.23 & 0.981 \\
 & PB, optimised & 0.108 & 1.39 & 0.947 & 0.151 & 1.20 & 0.984 \\
 & Classical, $\alpha=0$ & 0.143 & 2.99 & 0.826 & 0.103 & 4.26 & 0.829 \\
 & PB, $\alpha=0$ & 0.181 & 3.35 & 0.835 & 0.115 & 4.64 & 0.836 \\
\addlinespace
0.3 & Classical, optimised & 0.170 & 1.27 & 0.960 & 0.145 & 1.18 & 0.945 \\
 & PB, optimised & 0.190 & 1.38 & 0.977 & 0.165 & 1.21 & 0.964 \\
 & Classical, $\alpha=0$ & 0.192 & 2.81 & 0.834 & 0.097 & 3.76 & 0.808 \\
 & PB, $\alpha=0$ & 0.259 & 3.52 & 0.858 & 0.122 & 4.41 & 0.829 \\
\bottomrule
\end{tabular}

%% file: tables/tab_app_struct.tex
\begin{tabular}{ll cccc cc}
\toprule
Study & Question space & $\hat\phi$ & $\bar{\hat\nu}$ & $\bar{\hat\delta}$ (range) & $\bar N_{\mathrm{eff}}$ & $\Lambda$ & $p_\Lambda$ \\
\midrule
Political violence & full (type, region, period) & 1.00 & 0.95 & 1.93 (1.64--1.96) & 10.9 & 60.7 & $<$0.01 \\
 & quantity type & 0.25 & 0.21 & 1.73 (1.00--2.75) & 14.1 & 15.2 & $<$0.01 \\
 & region & 0.29 & 0.44 & 1.42 (1.00--1.76) & 15.2 & 63.7 & $<$0.01 \\
 & period & 0.17 & 0.26 & 1.43 (1.00--1.90) & 15.2 & 29.6 & $<$0.01 \\
\addlinespace
Terrorism and conflict & full (type, region, period) & 1.00 & 0.63 & 1.42 (1.00--1.71) & 13.0 & 20.3 & $<$0.01 \\
 & quantity type & 0.05 & 0.32 & 1.39 (1.00--2.38) & 14.2 & 25.9 & $<$0.01 \\
 & region & 0.23 & 0.32 & 1.24 (1.00--1.45) & 14.8 & 13.4 & 0.010 \\
 & period & 0.23 & 0.19 & 1.44 (1.00--2.36) & 13.8 & 13.0 & 0.015 \\
\addlinespace
Oil and growth & full (type, region, period) & 1.00 & 0.98 & 1.83 (1.83--1.83) & 10.9 & 10.1 & 0.025 \\
 & quantity type & 0.33 & 0.12 & 1.21 (1.02--1.46) & 16.7 & 2.2 & 0.365 \\
 & region & 0.31 & 0.41 & 1.58 (1.12--2.07) & 13.1 & 11.1 & 0.020 \\
 & period & 0.05 & 0.02 & 1.03 (1.00--1.17) & 19.5 & 0.5 & 0.745 \\
\bottomrule
\end{tabular}

%% file: tables/tab_app_sens.tex
\begin{tabular}{lll ccccc}
\toprule
Study & Expert & Specification & $\hat\phi$ & $\Lambda$ & $\bar{\hat\delta}$ & $\bar N_{\mathrm{eff}}$ & $C^{\mathrm{PB}}$ \\
\midrule
Political violence ($N=21$) & Exp7 & full, $3\phi\le3$ (baseline) & 1.00 & 60.7 & 1.93 & 10.9 & 0.047 \\
 &  & full, $4\phi\le3$ & 0.75 & 52.9 & 1.55 & 13.6 & 0.018 \\
 &  & full, $2\phi\le3$ & 1.50 & 64.1 & 2.60 & 8.2 & 0.100 \\
 &  & region only & 0.29 & 63.7 & 1.42 & 15.2 & 0.011 \\
\addlinespace
Terrorism and conflict ($N=18$) & Ex4 & full, $3\phi\le3$ (baseline) & 1.00 & 20.3 & 1.42 & 13.0 & 0.079 \\
 &  & full, $4\phi\le3$ & 0.75 & 16.8 & 1.24 & 14.7 & 0.043 \\
 &  & full, $2\phi\le3$ & 1.50 & 23.8 & 1.86 & 10.3 & 0.145 \\
 &  & quantity type only & 0.05 & 25.9 & 1.39 & 14.2 & 0.017 \\
\addlinespace
Oil and growth ($N=20$) & D & full, $3\phi\le3$ (baseline) & 1.00 & 10.1 & 1.83 & 10.9 & 0.107 \\
 &  & full, $4\phi\le3$ & 0.75 & 8.8 & 1.47 & 13.6 & 0.057 \\
 &  & full, $2\phi\le3$ & 1.50 & 11.1 & 2.57 & 7.8 & 0.212 \\
 &  & region only & 0.31 & 11.1 & 1.58 & 13.1 & 0.096 \\
\bottomrule
\end{tabular}

%% file: tables/tab_app_pv.tex
\begin{tabular}{l cccc cccc cc}
\toprule
Expert & $T_e$ & $\hat{\bar\delta}_e$ & $\hat N_{\mathrm{eff},e}$ & $\bar I_e$ & $C^{\mathrm{cl}}_e$ & $C^{\mathrm{RS2}}_e$ & $C^{\mathrm{PB}}_e$ & $C^{\mathrm{PBc}}_e$ & $w^{\mathrm{cl}}_e$ & $w^{\mathrm{PB}}_e$ \\
\midrule
Exp1 & 93.37 & 1.96 & 10.7 & 4.093 & $<$0.001 & $<$0.001 & $<$0.001 & 0.001 & 0.000 & 0.000 \\
Exp2 & 45.28 & 1.96 & 10.7 & 1.732 & $<$0.001 & $<$0.001 & $<$0.001 & 0.001 & 0.000 & 0.000 \\
Exp3 & 44.61 & 1.96 & 10.7 & 3.300 & $<$0.001 & $<$0.001 & $<$0.001 & 0.001 & 0.000 & 0.017 \\
Exp4 & 48.73 & 1.96 & 10.7 & 1.342 & $<$0.001 & $<$0.001 & $<$0.001 & 0.001 & 0.000 & 0.000 \\
Exp5 & 45.96 & 1.76 & 11.9 & 2.449 & $<$0.001 & $<$0.001 & $<$0.001 & 0.001 & 0.000 & 0.000 \\
Exp6 & 24.19 & 1.96 & 10.7 & 2.001 & $<$0.001 & 0.011 & 0.009 & 0.005 & 0.018 & 0.131 \\
Exp7 & 16.31 & 1.96 & 10.7 & 2.476 & $<$0.001 & 0.052 & 0.047 & 0.091 & 0.982 & 0.852 \\
Exp8 & 67.54 & 1.96 & 10.7 & 2.230 & $<$0.001 & $<$0.001 & $<$0.001 & 0.001 & 0.000 & 0.000 \\
Exp9 & 84.48 & 1.96 & 10.7 & 3.776 & $<$0.001 & $<$0.001 & $<$0.001 & 0.001 & 0.000 & 0.000 \\
Exp10 & 62.58 & 1.96 & 10.7 & 0.700 & $<$0.001 & $<$0.001 & $<$0.001 & 0.001 & 0.000 & 0.000 \\
Exp11 & 69.24 & 1.96 & 10.7 & 3.636 & $<$0.001 & $<$0.001 & $<$0.001 & 0.001 & 0.000 & 0.000 \\
Exp12 & 72.50 & 1.93 & 10.9 & 2.789 & $<$0.001 & $<$0.001 & $<$0.001 & 0.001 & 0.000 & 0.000 \\
Exp14 & 56.07 & 1.96 & 10.7 & 3.511 & $<$0.001 & $<$0.001 & $<$0.001 & 0.001 & 0.000 & 0.000 \\
Exp15 & 67.54 & 1.96 & 10.7 & 3.434 & $<$0.001 & $<$0.001 & $<$0.001 & 0.001 & 0.000 & 0.000 \\
Exp16 & 91.04 & 1.64 & 12.8 & 4.321 & $<$0.001 & $<$0.001 & $<$0.001 & 0.001 & 0.000 & 0.000 \\
\bottomrule
\end{tabular}

%% file: tables/tab_app_rav.tex
\begin{tabular}{l cccc cccc cc}
\toprule
Expert & $T_e$ & $\hat{\bar\delta}_e$ & $\hat N_{\mathrm{eff},e}$ & $\bar I_e$ & $C^{\mathrm{cl}}_e$ & $C^{\mathrm{RS2}}_e$ & $C^{\mathrm{PB}}_e$ & $C^{\mathrm{PBc}}_e$ & $w^{\mathrm{cl}}_e$ & $w^{\mathrm{PB}}_e$ \\
\midrule
Ex1 & 6.67 & 1.71 & 10.6 & 1.454 & 0.083 & 0.272 & 0.247 & 0.265 & 0.862 & 0.777 \\
Ex2 & 32.13 & 1.66 & 10.8 & 1.335 & $<$0.001 & $<$0.001 & 0.001 & 0.003 & 0.000 & 0.000 \\
Ex3 & 22.56 & 1.00 & 18.0 & 1.757 & $<$0.001 & $<$0.001 & $<$0.001 & 0.003 & 0.000 & 0.000 \\
Ex4 & 10.49 & 1.62 & 11.1 & 1.303 & 0.015 & 0.100 & 0.079 & 0.097 & 0.138 & 0.223 \\
Ex5 & 27.37 & 1.41 & 12.8 & 1.863 & $<$0.001 & $<$0.001 & $<$0.001 & 0.003 & 0.000 & 0.000 \\
Ex6 & 30.87 & 1.31 & 13.7 & 2.867 & $<$0.001 & $<$0.001 & $<$0.001 & 0.003 & 0.000 & 0.000 \\
Ex7 & 34.63 & 1.42 & 12.7 & 1.182 & $<$0.001 & $<$0.001 & $<$0.001 & 0.003 & 0.000 & 0.000 \\
Ex8 & 37.95 & 1.26 & 14.3 & 2.172 & $<$0.001 & $<$0.001 & $<$0.001 & 0.003 & 0.000 & 0.000 \\
\bottomrule
\end{tabular}

%% file: tables/tab_app_oil.tex
\begin{tabular}{l cccc cccc cc}
\toprule
Expert & $T_e$ & $\hat{\bar\delta}_e$ & $\hat N_{\mathrm{eff},e}$ & $\bar I_e$ & $C^{\mathrm{cl}}_e$ & $C^{\mathrm{RS2}}_e$ & $C^{\mathrm{PB}}_e$ & $C^{\mathrm{PBc}}_e$ & $w^{\mathrm{cl}}_e$ & $w^{\mathrm{PB}}_e$ \\
\midrule
B & 20.84 & 1.83 & 10.9 & 1.372 & $<$0.001 & 0.015 & 0.014 & 0.020 & 0.000 & 0.165 \\
C & 56.57 & 1.83 & 10.9 & 1.194 & $<$0.001 & $<$0.001 & $<$0.001 & 0.003 & 0.000 & 0.000 \\
D & 11.23 & 1.83 & 10.9 & 0.912 & 0.011 & 0.118 & 0.107 & 0.112 & 1.000 & 0.835 \\
E & 56.82 & 1.83 & 10.9 & 0.954 & $<$0.001 & $<$0.001 & $<$0.001 & 0.003 & 0.000 & 0.000 \\
F & 61.07 & 1.83 & 10.9 & 1.708 & $<$0.001 & $<$0.001 & $<$0.001 & 0.003 & 0.000 & 0.000 \\
H & 49.63 & 1.83 & 10.9 & 1.340 & $<$0.001 & $<$0.001 & $<$0.001 & 0.003 & 0.000 & 0.000 \\
\bottomrule
\end{tabular}

%% file: tables/tab_size_01.tex
\begin{tabular}{rr c ccccccc}
\toprule
$N$ & $\phi$ & $\bar\delta_N$ & $\widehat{\bar\delta}_N$ (med.) & Classical & RS1$^{\ast}$ & PB$^{\ast}$ & RS1 & RS2 & PB \\
\midrule
10 & 0.00 & 1.00 & 1.00 & 0.5 & 0.5 & 0.5 & 0.4 & 0.4 & 0.2 \\
 & 0.05 & 1.03 & 1.09 & 0.7 & 0.7 & 0.7 & 0.6 & 0.6 & 0.5 \\
 & 0.10 & 1.12 & 1.16 & 1.4 & 0.9 & 0.8 & 0.6 & 0.6 & 0.5 \\
 & 0.20 & 1.36 & 1.36 & \textbf{4.3} & 1.7 & 1.0 & 1.8 & 1.7 & 1.6 \\
 & 0.30 & 1.62 & 1.47 & \textbf{6.1} & 1.1 & 0.9 & 2.8 & 2.6 & 2.6 \\
\addlinespace
20 & 0.00 & 1.00 & 1.00 & 1.0 & 1.0 & 0.9 & 0.7 & 0.7 & 0.5 \\
 & 0.05 & 1.07 & 1.11 & 1.2 & 0.7 & 0.9 & 0.7 & 0.6 & 0.5 \\
 & 0.10 & 1.24 & 1.30 & \textbf{2.9} & 0.8 & 0.5 & 0.9 & 0.9 & 0.8 \\
 & 0.20 & 1.75 & 1.69 & \textbf{8.9} & 1.1 & 0.5 & 2.3 & 2.2 & 2.1 \\
 & 0.30 & 2.32 & 1.99 & \textbf{15.2} & 1.1 & 0.4 & 4.4 & 3.9 & 3.7 \\
\addlinespace
30 & 0.00 & 1.00 & 1.00 & 1.3 & 1.3 & 1.0 & 0.8 & 0.8 & 0.6 \\
 & 0.05 & 1.11 & 1.14 & \textbf{1.6} & 0.9 & 0.9 & 0.8 & 0.8 & 0.6 \\
 & 0.10 & 1.38 & 1.39 & \textbf{4.1} & 1.3 & 0.9 & 1.7 & 1.6 & 1.4 \\
 & 0.20 & 2.16 & 2.02 & \textbf{14.0} & 1.5 & 0.8 & 3.2 & 2.9 & 2.7 \\
 & 0.30 & 3.02 & 2.59 & \textbf{22.4} & 1.3 & 0.8 & 5.5 & 5.0 & 4.7 \\
\addlinespace
50 & 0.00 & 1.00 & 1.00 & 1.2 & 1.2 & 1.0 & 1.1 & 1.1 & 0.8 \\
 & 0.05 & 1.18 & 1.21 & \textbf{2.6} & 0.8 & 0.6 & 1.0 & 1.0 & 0.7 \\
 & 0.10 & 1.64 & 1.62 & \textbf{8.2} & 1.5 & 1.0 & 2.0 & 1.8 & 1.6 \\
 & 0.20 & 2.95 & 2.79 & \textbf{23.9} & 1.3 & 0.4 & 3.4 & 2.8 & 2.3 \\
 & 0.30 & 4.40 & 3.79 & \textbf{41.1} & 1.7 & 0.3 & 7.8 & 6.0 & 5.4 \\
\addlinespace
100 & 0.00 & 1.00 & 1.00 & \textbf{1.8} & 1.8 & 0.9 & 1.4 & 1.4 & 0.8 \\
 & 0.05 & 1.36 & 1.35 & \textbf{5.9} & 2.3 & 1.6 & 2.1 & 1.9 & 1.4 \\
 & 0.10 & 2.28 & 2.27 & \textbf{16.8} & 1.9 & 0.6 & 2.3 & 1.8 & 1.0 \\
 & 0.20 & 4.96 & 4.47 & \textbf{47.9} & 2.0 & 0.3 & 7.2 & 6.1 & 5.4 \\
 & 0.30 & 7.86 & 7.35 & \textbf{55.7} & 1.7 & 1.0 & 5.5 & 4.1 & 3.4 \\
\bottomrule
\end{tabular}